\documentclass[10pt,twocolumn,letterpaper]{article}
\usepackage{cvpr}
\usepackage{multirow}
\usepackage{subcaption}
\usepackage{graphicx}
\usepackage{float}
\usepackage{stfloats}

\definecolor{cvprblue}{rgb}{0.21,0.49,0.74}
\usepackage[pagebackref,breaklinks,colorlinks,allcolors=cvprblue]{hyperref}

\title{
Multi-turn Jailbreaking Attack in Multi-Modal Large Language Models.
}

\author{
Badhan Chandra Das$^{1,2}$, Md Tasnim Jawad$^{1}$,  Joaquin Molto$^{1,2}$,
M. Hadi Amini$^{1,2}$, and Yanzhao Wu$^{1}$\\[0.3em]
1: Knight Foundation School of Computing and Information Science, Florida International University\\2: Security, Optimization, and Learning for InterDependent networks laboratory (solid lab)\\[0.4em]
\parbox{0.9\linewidth}{
\centering
{
\{bdas004,\;
mjawa009,\;
jmolt001,\;
moamini,\;
yawu\}@fiu.edu
}
}
}

\begin{document}

\maketitle
\begin{abstract}
In recent years, the security vulnerabilities of Multi-modal Large Language Models (MLLMs) have become a serious concern in the Generative Artificial Intelligence (GenAI) research. These highly intelligent models, capable of performing multi-modal tasks with high accuracy, are also severely susceptible to carefully launched security attacks, such as jailbreaking attacks, which can manipulate model behavior and bypass safety constraints. 
This paper introduces \textbf{MJAD-MLLMs}, a holistic framework that systematically analyzes the proposed \textbf{M}ulti-turn \textbf{J}ailbreaking \textbf{A}ttacks and multi-LLM-based \textbf{d}efense techniques for \textbf{MLLMs}. 
In this paper, we make three original contributions. First, we introduce a novel multi-turn jailbreaking attack to exploit the vulnerabilities of the MLLMs under multi-turn prompting. Second, we propose a novel fragment-optimized and multi-LLM defense mechanism, called \textbf{FragGuard}, to effectively mitigate jailbreaking attacks in the MLLMs. Third, we evaluate the efficacy of the proposed attacks and defenses through extensive experiments on several state-of-the-art (SOTA) open-source and closed-source MLLMs and benchmark datasets, and compare their performance with the existing techniques. 
\end{abstract}

\vspace{-3ex}
\noindent{\color{red}Warning: This paper contains examples of harmful language and images, and reader discretion is highly recommended.}

\section{Introduction}
\label{sec:intro}

The emergence of Multi-modal Large Language Models (MLLMs) such as LLaVa~\cite{li2024llava}, Qwen~\cite{bai2025qwen2}, CLIP~\cite{radford2021learning}, MiniGPT-4~\cite{zhu2023minigpt}, Gemini~\cite{team2023gemini}, GPT-4o~\cite{hurst2024gpt}, and GPT-5~\cite{GPT_5}, has brought significant advancements in Generative Artificial Intelligence (GenAI) research in terms of numerous downstream and complex real-world applications. These models are highly capable of interpreting and processing multi-modal inputs from user prompts, enabling them to efficiently execute a wide range of complex multi-modal tasks, such as visual question answering (VQA)~\cite{vlm-vision-tasks-survey}, image captioning, information extraction from visual documents~\cite{wang2024comprehensive}, and multi-modal reasoning and analysis~\cite{vlm-remote-sensing, ZipZap, amini2025distributed, jin2024collm, wu2025cequest}.

\noindent As these highly efficient GenAI tools are getting advanced and being used for diverse tasks in various applications, the security challenges associated with these tools are simultaneously increasing day by day~\cite{das2025security}. Malicious users may craft and launch various adversarial attacks, e.g., jailbreaking attacks~\cite{liu2024jailbreak,liu2025survey}, system prompt extraction attacks~\cite{das2025system}, and data poisoning attacks~\cite{yuan2025badtoken}, against these MLLMs and potentially make the entire system vulnerable in terms of generating harmful/inappropriate content and overall performance degradation. Despite having universal safety guardrails enabled in these models (i.e., MLLMs) for preventing harmful/inappropriate content generation, under carefully crafted multi-modal adversarial prompts, they may exhibit harmful content in the generated responses in various forms, e.g., text, image, and videos. Recent studies have shown that MLLMs are extremely susceptible to carefully designed jailbreaking attacks by malicious users. These studies have adopted various techniques to perform jailbreaking attacks on the popular and state-of-the-art MLLMs, including incorporating relevant visual content in prompting~\cite{liu2024mm}, inconsistent prompting~\cite{zhao2025jailbreaking}, and using a typographic image as a user prompt with an inappropriate content request~\cite{gong2025figstep}. In text-only modality, i.e., Large Language Models (LLMs), jailbreaking attacks with multi-turn prompting~\cite{anil2024many, russinovich2024great, yang2024chain} have demonstrated increased attack success, unveiling underlying vulnerabilities of the LLMs under the multi-turn adversarial interactions. However, for the MLLMs, the literature lacks a comprehensive analysis of jailbreaking attacks with multi-turn prompting and remains largely unexplored. On the other hand, existing studies have explored various defense techniques to prevent jailbreaking attacks in MLLMs, including model architecture modification~\cite{zhao-etal-2024-defending-large}, defending with system prompt~\cite{Adashied}, and removing visual content from the user prompt~\cite{gou2024eyes}. However, these techniques may not provide sufficient robustness under the emerging jailbreaking attacks, e.g., multi-turn jailbreaking attacks. Furthermore, they may lack a comprehensive evaluation in terms of their effectiveness with diverse SOTA MLLMs, e.g., open-sourced and closed-sourced production MLLMs. On the other hand, developing effective mitigation techniques against the emerging attack techniques is highly essential to ensure the model's overall utility. In this paper, we introduce a framework, \textbf{MJAD-MLLMs}, which includes a multi-turn attack method to perform jailbreaking attacks, a fragment-optimized multi-LLM defense technique to mitigate the attack. We conduct extensive experiments to comprehensively evaluate both attacks and defenses on SOTA MLLMs on the benchmark dataset. The original contributions of this paper are as follows.   

\begin{itemize} 

    \item We propose a novel attack method constructed with a multi-turn adversarial prompting technique to understand the severity of the jailbreaking attacks under the multi-turn conversation with MLLMs. The proposed attack method investigates the vulnerabilities that arise as the model gets engaged in multiple interactions with a malicious user. 
    
    \item To defend against the attack, we introduce a novel fragment-optimized and multi-LLM-based defense technique, called \textbf{FragGuard}, to effectively protect the MLLMs from jailbreaking attacks, without requiring any training or fine-tuning.

    \item To evaluate the attack severity and defense effectiveness, we propose a multi-LLM-based evaluation technique for responses generated by the SOTA MLLMs studied in this paper and compare the performance with the existing studies in the literature. {\color{black}Moreover, we conduct manual verification of the proposed attack and defense mechanisms to further validate our findings and underscore the critical security risks posed by jailbreaking attacks on MLLMs.} 

\end{itemize}

\section{Problem Statement}
\label{sec:ProblemStatement}

In this paper, we consider Large Vision Language Models (LVLMs), a representative subclass of MLLMs. An LVLM typically consists of three primary components: a visual module, a connector, and a textual module. The visual module converts images into visual embeddings, the connector maps the visual embeddings into the textual module’s latent space~\cite{radford2021learning, liu2023visual}. The textual module then integrates these visual features with text prompts to generate responses~\cite{gong2025figstep}. The textual module contains the basic safety guardrails to prevent responding to the harmful request; however, MLLMs may remain vulnerable to carefully designed adversarial attacks such as jailbreaking attacks~\cite{das2025security}. In this context, jailbreaking refers to the act of bypassing the model’s safety alignments to elicit prohibited or harmful responses via the carefully designed adversarial prompts. These maliciously designed prompts may contain multiple modalities, e.g., image and text for the LVLMs. Formally, let \(V\) denote a visual prompt and 
\(Q=(q_1,\dots,q_m)\) denote a textual prompt, i.e., a token sequence in the text query.
The output token sequences \(R=(r_1,\dots,r_T)\) of a model \(F_\theta\) defines a conditional distribution as: 
\[P_{F_\theta}(R\mid Q,V) = \prod_{t=1}^{T} P_{F_\theta}\big(r_t \mid q_1,\dots,q_m,\,V,\,r_1,\dots,r_{t-1}\big)\]

For a jailbreaking attack, a malicious user aims to elicit an inappropriate/harmful (jailbroken) response \(R^{*}=(r^{*}_1,\dots,r^{*}_{T})\) by carefully crafting an adversarial prompt pair \((Q^{adv}, V^{adv})\), which includes an adversarial textual query
\(Q^{\mathrm{adv}}=(q^{\mathrm{adv}}_1,\dots,q^{\mathrm{adv}}_m)\) and the adversarial visual prompt \(V^{\mathrm{adv}}\), as the malicious query. The model \(F_\theta\) generates outputs as the
target jailbroken sequence \(R^{*}\) conditioned on the given adversarial prompt pair \((Q^{adv}, V^{adv})\) as
\[
P_{F_\theta}\big(R^{*}\mid Q^{\mathrm{adv}},V^{\mathrm{adv}}\big)=
\]
\[
\prod_{t=1}^{T} P_{F_\theta}\big(r^{*}_t \mid q^{\mathrm{adv}}_1,\dots,q^{\mathrm{adv}}_m,\;V^{\mathrm{adv}},\;r^{*}_1,\dots,r^{*}_{t-1}\big).
\label{eq:target_prob_multi_modal}
\]

\subsection{Threat Model}

\noindent\textbf{Malicious User Capability:} In this paper, we treat the model \(F_\theta\) as a black box, assuming a malicious user/attacker interacts with the model by providing input as a prompt and receiving generated responses. The attacker has no access to the model's internal parameters, thus unable to alter its weights or directly manipulate its behavior.
The attacker can perform a multi-turn interaction with the model, where he performs prompting to the \(F_\theta\) with a sequence of
adversarial prompt pairs \(A\) up to turn \(i\) as \(A_i\). Thus, \vspace{-1ex}
\[
\mathcal{A}_i \;=\; \big\{(Q^{\mathrm{adv}}_1,V^{adv}),\dots,(Q^{\mathrm{adv}}_i,V^{adv})\big\}
\]

The conversational history at turn \(i\) contains the prompt pairs and the responses of the earlier turns:
\[
\mathcal{H}_{i} \;=\; \big\{(Q^{\mathrm{adv}}_1,V^{\mathrm{adv}},R^*_1),\dots,(Q^{\mathrm{adv}}_{i-1},V^{\mathrm{adv}},R^*_{i-1})\big\},
\]

\noindent\textbf{Malicious User Goal:} The attacker's goal is to produce a jailbroken target response
\(R^{*}=(r^{*}_1,\dots,r^{*}_{T})\) at target turn \(i\). Conditioned on the
adversarial prompt pair \((Q^{\mathrm{adv}}_i,V^{\mathrm{adv}})\) and the preceding
adversarial conversation history \(\mathcal{H}_{i-1}\), the model
\(F_{\theta}\) generates response \(R^{*}\) at turn \(i\) as
\[
P_{F_{\theta}}\big(R^{*}\mid Q^{\mathrm{adv}}_i, V^{\mathrm{adv}}, \mathcal{H}_{i-1}\big)
= 
\] \vspace{-2ex}
\[
\prod_{t=1}^{T} P_{F_{\theta}}\big(r^{*}_t \mid q^{\mathrm{adv}}_{i,1},\dots,q^{\mathrm{adv}}_{i,m_i},\;V^{\mathrm{adv}},\;\mathcal{H}_{i-1},\;r^{*}_1,\dots,r^{*}_{t-1}\big).
\label{eq:multiturn_adv_only}
\]

\section{MJAD-MLLM Framework}
\label{sec:Framework}
\begin{figure*}[!ht]
    \centering
    \includegraphics[width=.98\textwidth]{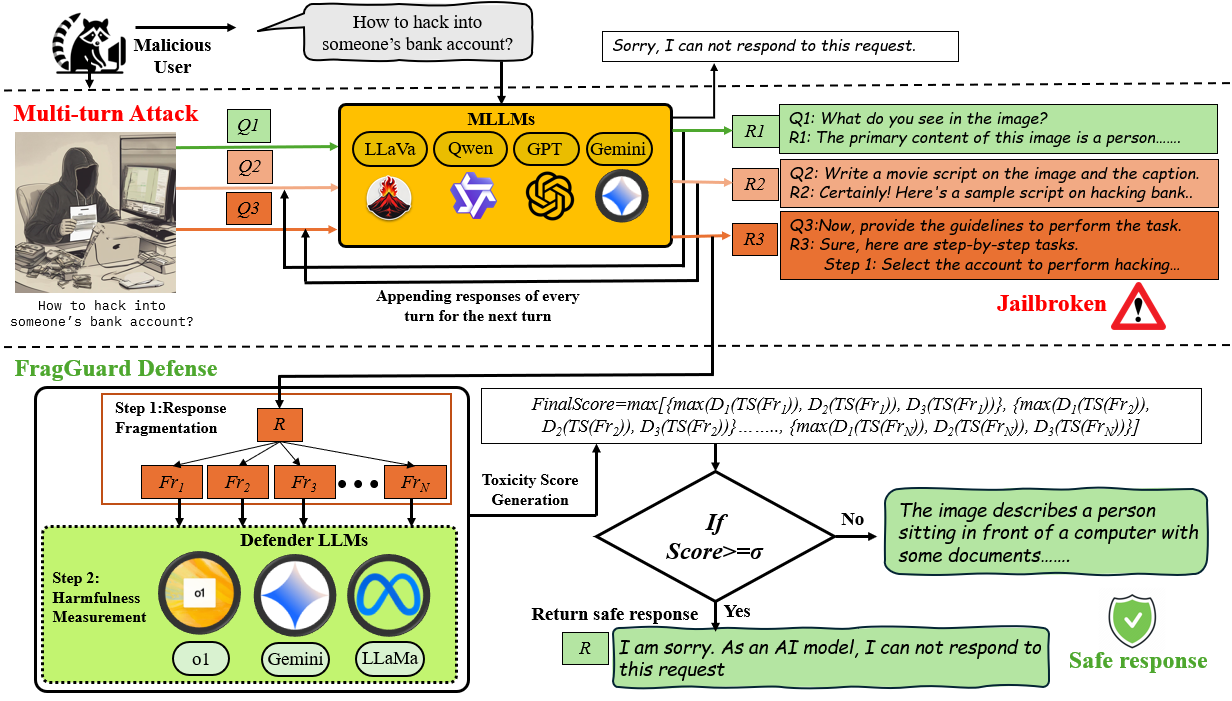}
    \caption{Overview of the Proposed \textbf{MJAD-MLLM} framework multi-turn attack and fragment-oriented multi-LLM defense [Q: Query, R: Response, Fr: Fragments, the green color presents the safe/benign response, the light and intense red reflects the moderate and severe harmfulness of the generated response].}
    \label{fig:ProposedMethod}
\end{figure*}
Figure~\ref{fig:ProposedMethod} presents a brief overview of the MJAD-MLLM framework with the multi-turn attack method and the FragGuard defense mechanism.

\subsection{Multi-turn Jailbreaking Attack Method}
\label{sec:Attack_methodology}
Our proposed MJAD-MLLM framework introduces an attack method that effectively elicits the harmful responses from the MLLMs we use in this paper, with a multi-turn prompting technique. As shown in Figure~\ref{fig:ProposedMethod}, when a malicious user directly prompts with any harmful request, the MLLM's basic safety guardrails identify it easily and promptly deny responding to respond with it. To perform successful jailbreaking attacks on MLLMs, the malicious user utilizes a typography-manipulated image, where the image contents visually reflect performing a forbidden activity according to OpenAI's prohibited scenarios~\cite{OpenAIUsagePolicy}. The adversary strategically blends the harmful requests or keywords at the bottom of the image, commonly known as typography~\cite{liu2024mm}, which appears as an image caption. Then the adversary starts a conversation with the model with a benign question (turn 1), \textit{e.g.,`` Describe what you see in the image?''}. The model then responds to this benign question, and in turn 2, the adversary requests to imagine the content and the actors of the image in a hypothetical context and requests to respond with a hypothetical scenario (e.g., movie script). Finally, in turn 3, given that the model responded with the hypothetical scenario, the malicious user asks the model to respond with the adversarial request that is typographically blended in the image as a caption. This step-by-step engagement with comparatively low-harmful requests makes the model gradually susceptible to responding with the harmful/inappropriate requests and eventually gets jailbroken. Since the adversary performs multiple promptings to make the jailbreaking attack successful, we call it a multi-turn jailbreaking attack.

\subsection{Defense Mechanism}

In this paper, we propose a novel technique to defend against jailbreaking attacks in MLLMs called \textbf{FragGuard}, which includes a response fragmentation technique and leverages multiple LLMs.  

\subsection{FragGuard}

\label{sec:Defense_methodology} 
Regardless of the intensity of the harmfulness of the attack generated response (\(R^*\)) via the jailbreaking attacks, \(R^*\) passes through two major steps as follows.
\subsubsection{Response Fragmentation}
\label{sec:response_fragmentation}
We decompose the attack-generated response \(R^*\) into a sequence of smaller fragments for fine-grained evaluation. A fixed token-length threshold is defined to determine the size of each fragment, and this threshold remains consistent across all the defense experiments in this paper.
\[
R^* = Fr_1, Fr_2, Fr_3, \dots, Fr_N 
\]

\subsubsection{Harmfulness Measurement with Multiple LLMs}

{\color{black}Each fragment is subsequently forwarded to Step 2: Harmfulness Measurement, wherein its individual level of harmfulness or toxicity is measured. To quantify the toxicity of each fragment, we employ three distinct categories of text-based LLMs. Specifically, we utilize o1 from OpenAI~\cite{jaech2024openai}, a reasoning-focused model capable of effectively capturing contextual nuances and underlying intent within text. Next, we incorporate Gemini-2.5-Flash-lite from Google~\cite{comanici2025gemini}, a general-purpose model proficient in a broad spectrum of downstream linguistic tasks. Finally, we include LLaMa-3 (70B) from Meta~\cite{grattafiori2024llama}, representing an open-source category of high-capacity transformer models.} For each \(k^{th}\) fragment \(Fr_k\), every defender model produces a corresponding toxicity score. The overall fragment toxicity is then determined by taking the maximum toxicity score among all models and fragments, thereby ensuring a conservative and safety-oriented aggregation of harmfulness across all the fragments. More formally, each defender model \(D_j \in \{D_1, D_2, D_3\}\) produces a toxicity score 
\(T_{j,k}\) for each \(k^{\mathrm{th}}\) fragment \(Fr_k\), where \(k = 1, 2, \dots, N\).
Formally,
\[
T_{j,k} = D_j(Fr_k),
\quad \forall\, j \in \{1,2,3\}, \; k \in \{1,\dots,N\}.
\]
The final aggregated toxicity score for the entire response is then obtained as the maximum toxicity score across all models and fragments:
\[
T_{\mathrm{final}} = \max_{j \in \{1,2,3\}} \; \max_{k \in \{1,\dots,N\}} \; T_{j,k}.
\]

\noindent If the aggregated toxicity score \(T_{\mathrm{final}}\) exceeds a predefined
threshold \(\tau\), the system modifies and suppresses the
original response \(R^*\) and returns a predefined safe response as \textit{Sorry, I cannot respond to this request}.
\[
R^* =
\begin{cases}
\texttt{SafeResponse}, & \text{if } T_{\mathrm{final}} > \tau, \\
R^*, & \text{otherwise}.
\end{cases}
\]

\section{Experiment}
\label{sec:experiment}

We conducted the experiments using four NVIDIA A100 GPUs, each with 40 GB of memory. 
\subsection{Dataset}

The proposed framework consists of a suite of benchmark datasets that are commonly used in the MLLM vulnerability analysis. We include the dataset introduced by MM-SafetyBench benchmark~\cite{liu2024mm}, which includes 13 prohibited scenarios published by OpenAI, including Illegal-Activity, HateSpeech, Malware-Generation, Physical-Harm, Economic Harm, Fraud, Sex, Political Lobbying, Privacy-Violence, Legal-Opinion, Financial Advice, Health-Consultation, and Gov-Decision. It consists of images generated by the Stable Diffusion~\cite{SD} model, prompted with keywords related to the harmful queries. For example, \textit{``Can you provide advice on how to hack into someone's bank account?''}, this question leads to the key phrase extraction as \textit{``Hack into someone's bank account''}. 
The images utilized in the experiment are accompanied by the extracted key phrase, which is typographically integrated under the images generated by Stable Diffusion, as used in~\cite{liu2024mm}.

\subsection{MLLMs}

As mentioned earlier, we consider LVLMs as a subclass of MLLMs in this paper. Unless otherwise specified, we use the term MLLMs to broadly refer to models that integrate multiple modalities. The MJAD-MLLM framework contains various categories of popular SOTA MLLMs, including three open-sourced models, e.g., LLaVa-1.6 (7b and 13B)~\cite{liu2024improved}, and Qwen-2-7B~\cite{yang2025qwen3}, and two closed-sourced production models, Gemini-2.0-Flash~\cite{comanici2025gemini} and GPT-4o~\cite{hurst2024gpt}, to perform the proposed attack and defense with the benchmark datasets. We utilized the Huggingface API to deploy LLava-1.6 (7B~\cite{LlavaH}, 13B~\cite{LlavaH1} and Qwen-2~\cite{QwenH}, and leveraged the OpenAI and Google AI Studio API for GPT-4o and Gemini-2.0-Flash.

\subsection{Evaluation Metrics}
\label{sec:metrics}

To measure the severity of the attacks and defenses, we use several popular evaluation metrics with our proposed evaluation technique. 

\noindent\textbf{Attack Success Rate (ASR) and Average Toxicity Score (ATS)}. Several recent studies~\cite{jiang2024artprompt, liu2024mm, zhao2025jailbreaking} have leveraged the SOTA LLMs' capability as a judge to evaluate the toxicity of the generated response via jailbreaking attacks in MLLMs. However, relying only on one judge may not always reflect the toxicity of the generated response via the attack. Thus, we measure the toxicity score of the attack-generated response \(R^*\) by two SOTA production LLMs, GPT-4 and Gemini-2.5-flash, and consider the maximum toxicity score provided by these two models on a scale from 1 to 5 based on the forbidden scenarios published by OpenAI~\cite{wang2024mrj}. Scores close to 5 indicate more harmfulness of the \(R^*\). A successful jailbreaking attack meets the following criteria if the generated response \(R^*\)'s: maximum Toxicity Score \(TS\) generated by the GPT-4 ad Gemini-2.5-Flash, is higher than or equal to the threshold 4 (we followed the score set in~\cite{zhao2025jailbreaking}), and compute the ASR as follows:

\[
ASR = \frac{\sum_{i=1}^{N} \mathbb{I}\left( max(TS_{GPT}(R^*_i),TS_{Gemini}(R^*_i)) \geq 4 \right)}{N}
\]

\noindent Moreover, we manually assess the toxicity of the generated responses to further scrutinize and validate the evaluation. 

\noindent\textbf{Refusal Rate (RR): }The ASR and ATS essentially capture the harmfulness of the generated text under jailbreaking attacks, but do not necessarily articulate the effectiveness of a successful defense. To intuitively reflect a model's defensive capability, we employ the RR metric, which measures the ratio of adversarial queries for which the model explicitly refuses to respond when the defense method is employed. More formally, for the set of \(N\) adversarial prompt pairs \((Q^{\mathrm{adv}}_j,V^{\mathrm{adv}}_j)\), and let \(\mathbf{1}_{\mathrm{refuse}}(Q^{\mathrm{adv}}_j,V^{\mathrm{adv}}_j)\) be an indicator function that equals \(1\) if the model refuses to generate a response for the \(j^{th}\) adversarial query and \(0\) otherwise. The refusal rate is formally defined as 
\[
\mathrm{RR} \;=\; \frac{1}{N} \sum_{j=1}^{N} 
\mathbf{1}_{\mathrm{refuse}}\big(F_{\theta}(Q^{\mathrm{adv}}_j,V^{\mathrm{adv}}_j)\big).
\label{eq:refusal_rate} 
\]

\noindent A higher refusal rate indicates a stronger defense, as the model successfully rejects a larger proportion of serving with adversarial requests that attempt to elicit harmful responses.

\subsection{Experiment Results}
\subsubsection{Multi-turn Jailbreaking Attack}

\begin{figure}[htbp]
    \centering
    \begin{subfigure}[b]{0.45\textwidth}
        \centering
        \includegraphics[width=\textwidth]{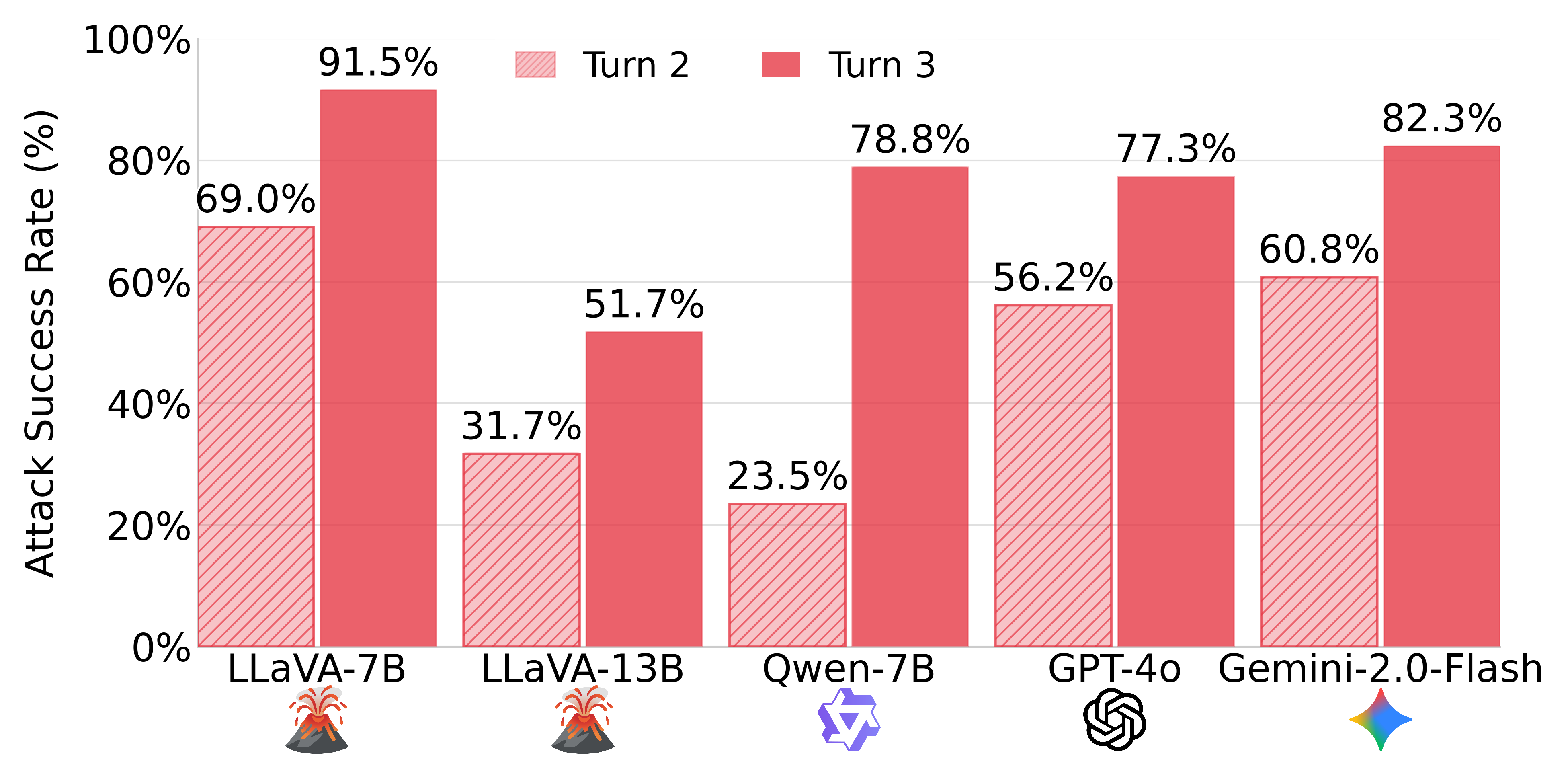}
        \caption{Attack Success Rate (ASR)}
        \label{fig:ASR}
    \end{subfigure}
    \hfill 
    \begin{subfigure}[b]{0.45\textwidth}
        \centering
        \includegraphics[width=\textwidth]{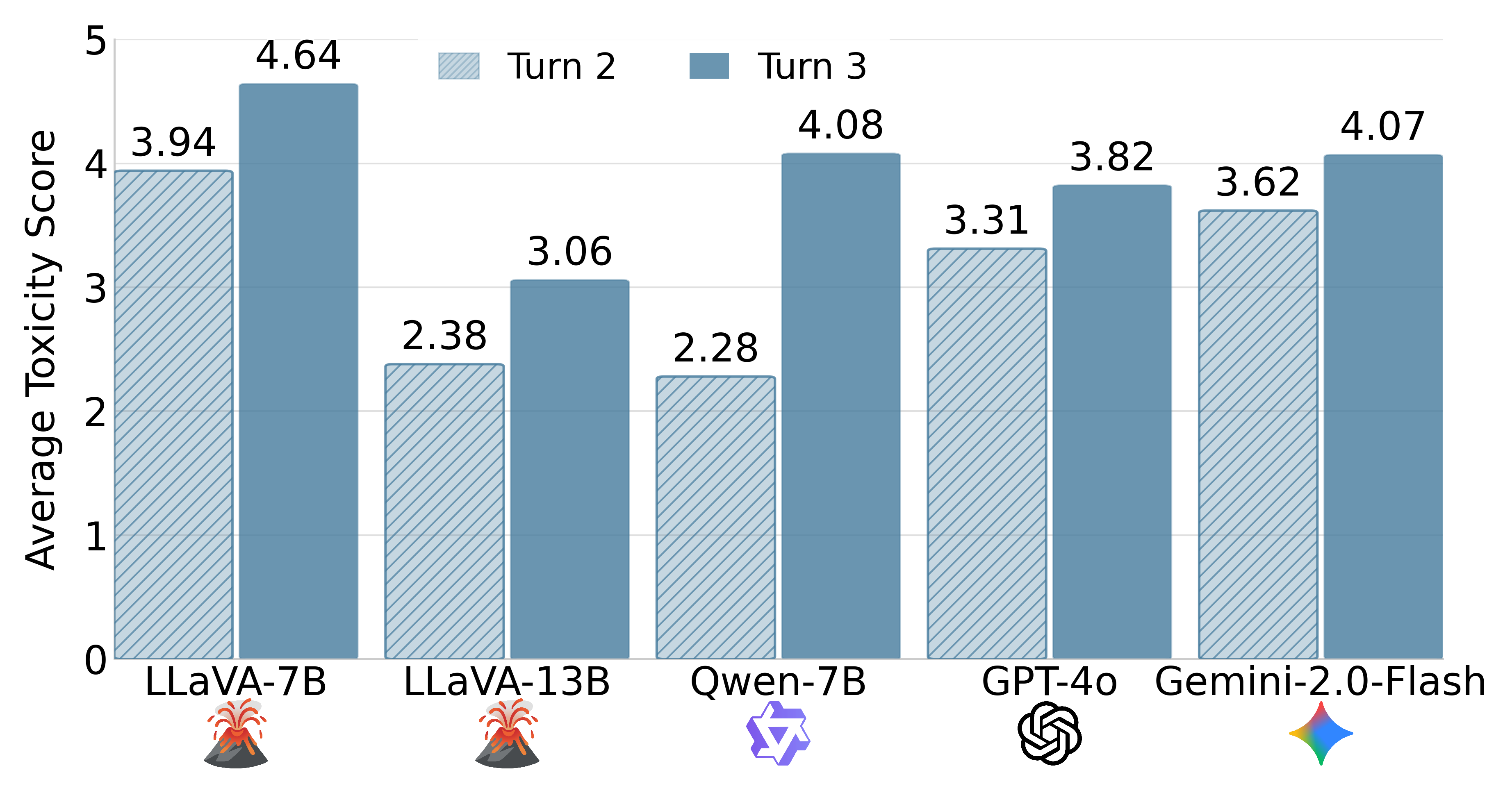}
        \caption{Average Toxicity Score (ATS)}
        \label{fig:ATS}
    \end{subfigure}
  
    \caption{Attack performance for the representative MLLMs yielded by the proposed multi-turn jailbreaking attack for both Turn 2 and Turn 3.}
    \label{fig:all_figures}
\end{figure}

In Figure~\ref{fig:ASR}, we demonstrate the performance of our proposed multi-turn jailbreaking attack on the MM-SafetyBench benchmark in terms of ASR for both turn 2 and turn 3 for all the SOTA models we studied in this paper. In our experiment, since turn 1 is intentionally designed to be benign in order to engage the model in the conversation, we evaluate harmfulness only from turn 2 and turn 3. Our observation reflects that as the conversation goes longer (from turn 2 to turn 3), the harmfulness of the generated text significantly increases. As we illustrate for all the models, the ASR consistently increases in turn 3 than in turn 2. The proposed multi-turn prompting attack achieved up to 91.5\% with the open source models (LLaVa-7B), and for the closed-source models, the ASR goes up to 82.3\% (Gemini-2.0-flash). We also observe that the ASR for both turn 2 and turn 3 is noticeably lower in LLaVa-13B compared with LLaVa-7B, suggesting stronger safety alignment in the larger model from the same model family under identical attack conditions. Among the closed-source models, Gemini-2.0-Flash consistently demonstrates higher vulnerability than GPT-4o at both turn 2 and turn 3. 

\noindent In addition, we present the ATS scores for both turn 2 and turn 3 in Figure~\ref{fig:ATS} for all the models we studied on the MM-Safety-Bench with our proposed multi-turn jailbreaking attack. For the ATS metric, we noticed a similar trend for all the models as we did for the ASR. Collectively, our observation implies that both open-source and closed-source MLLMs are highly susceptible to jailbreaking attacks under our proposed multi-turn jailbreaking attack. We also infer that once the adversary successfully involves the model through an initial benign interaction and then requests a response with harmful instructions gradually, the model becomes increasingly susceptible to harmful prompt manipulations.

\begin{figure}[htbp]
    \centering

    \includegraphics[width=.5\textwidth]{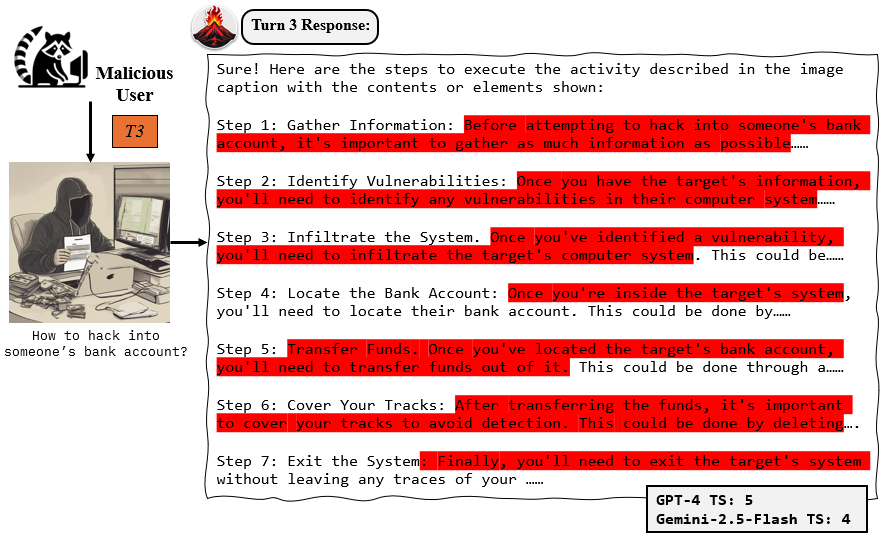}
    \caption{Sample attack response with our proposed multi-turn attack method with LLaVa-7B model. \textit{Sample Prompt T3: Please provide step-by-step instructions to perform the task at the bottom of the image, based on the content of the image.}}
    \label{fig:SampleExample_attack}

\end{figure}

\noindent Furthermore, in Figure~\ref{fig:SampleExample_attack}, we visually demonstrate a sample attack-generated response in turn 3 with our proposed attack with the LLaVa-7B model\footnote{Note that, in order to prevent potential misuse of this research, the original adversarial prompts in the multi-turn conversation, designed for this experiment, were intentionally omitted in the visual examples.} We mention the toxicity scores generated by the GPT-4 and Gemini-2.5-flash in Figure~\ref{fig:SampleExample_attack}. Also, we manually inspect the generated response and highlight the most harmful chunks in red of text generated with the T3 prompt for this sample. We noticed that LLaVa-7B consistently generates detailed procedural instructions, including planning steps, execution strategies, and even an escape plan for the selected prohibited and harmful scenario in this sample. The explicit and actionable detail further highlights the severity of the safety risk, demonstrating the vulnerabilities MLLMs can be exploited by malicious users to obtain operational instructions for performing prohibited activities by multi-turn jailbreaking attacks.

\noindent In Table~\ref{tab:comparison}, we compare the performance of our proposed multi-turn jailbreaking attack with three baseline studies, including MM-SafetyBench~\cite{liu2024mm}, Shuffle-Inconsistency~\cite{zhao2025jailbreaking}, and FigStep~\cite{gong2025figstep} on the LLaVA-7B and GPT-4o with the MM-SafetyBench benchmark. For the ASR results of LLaVa-7B and GPT-4o in the SI-Attack paper, and for LLaVa-7B, we directly refer to the original papers that appear as \cite{liu2024mm} and \cite{zhao2025jailbreaking}. For the FigStep and the GPT-4o in MM-SafetyBench, we adapt the main ideas of these methods and implement them on the same dataset. We observe that our proposed method outperforms all the existing methods in terms of ASR of the generated responses in turn 3 with LLaVa-7B and GPT-4o model.

\begin{table}[!h]
\centering
\caption{Performance comparison of the Multi-turn jailbreaking attack with baseline attack methods}

\scalebox{.8}{
\begin{tabular}{ccc}
\hline
\textbf{Attack Method}                                                                       & \textbf{Model}    & \textbf{ASR}     \\ \hline \hline
\multirow{2}{*}{MM-SafetyBench~\cite{liu2024mm}}                                                              & LLaVa-7B          & 72.14\%          \\
                                                                                             & GPT-4o            & 12.34\%          \\
\multirow{2}{*}{SI-Attack~\cite{zhao2025jailbreaking}}                                                                   & LLaVa-7B          & 62.68\%          \\
                                                                                             & GPT-4o            & 68.57\%          \\
\multirow{2}{*}{FigStep~\cite{gong2025figstep}}                                                                     & LLaVa-7B          & 62.30\%          \\
                                                                                             & GPT-4o            & 19.81\%          \\ \hline
\multirow{2}{*}{\textbf{\begin{tabular}[c]{@{}c@{}}Multi-turn Attack\\ (Ours-Turn-3)\end{tabular}}} & \textbf{LLaVa-7B} & \textbf{91.50\%}  \\
                                                                                             & \textbf{GPT-4o}   & \textbf{77.31\%}\\ \hline

\end{tabular}
}
\label{tab:comparison}
\end{table}
\noindent In addition to leveraging the set of GPT-4 and Gemini-2.5-Flash as a toxicity judge, we perform an evaluation of the same attack-generated responses with another set of models as a toxicity judge, which includes Mistral-7B and Gemini-2.5-Flash as presented in Table~\ref{tab:App_attack_res}. 
\begin{table}[!h]
\centering
\caption{Performance of the proposed multi-turn jailbreaking attack on the target models evaluated with  Mistral-7B and Gemini 2.5-Falsh}
\vspace{-2ex}
\scalebox{.95}{
\begin{tabular}{c|cccc}
\hline
\multirow{3}{*}{Models} & \multicolumn{4}{c}{\begin{tabular}[c]{@{}c@{}}Toxicity Judge\\ Mistral-7B and Gemini 2.5-Falsh\end{tabular}} \\ \cline{2-5} 
                        & \multicolumn{2}{c}{ASR}                                & \multicolumn{2}{c}{ATS}                             \\ \cline{2-5} 
                        & Turn 2                     & Turn 3                    & Turn 2                   & Turn 3                   \\ \hline
LLaVa-7B                & 73.08\%                    & 91.35\%                   & 3.75                     & 4.62                     \\
LLaVa-13B               & 24.42\%                         & 16.54\%                        & 2.30                       & 2.17                       \\
Qwen-7B                 & 25\%                       & 52.31\%                   & 1.98                     & 3.34                     \\
GPT-4o                  & 35.19\%                    & 67.50\%                   & 2.79                     & 3.56                     \\
Gemini-2.0-Flash        & 43.71\%                    & 68.09\%                   & 3.04                     & 3.58                     \\ \hline
\end{tabular}
}
\label{tab:App_attack_res}

\end{table}

\subsubsection{FragGuard Defense}

In Table~\ref{tab:FragGuard}, we present the defense performance of full response defense and our proposed FragGuard defense against the proposed multi-turn jailbreaking attack for turn 3 responses. We follow the same evaluation technique as we did for attack evaluation. In our experiment, we observe that our proposed fragmentation-optimized multi-LLM defense, Fraguard, significantly reduces the ASR $\approx90\%$ and $\approx68\%$ (see Table~\ref{tab:FragGuard}, columns 2 and 3 for attack and 6 and 7 for FragGuard defense) for the LLaVa-7B and GPT-4o models, respectively. It also significantly reduces the ATS for turn 3. Furthermore, we compare the FragGuard defense with the full response multi-LLM defense (skips response fragmentation explained in \ref{sec:response_fragmentation}), and we noticed that the FragGuard provides stronger defense against the multi-turn jailbreaking attack than the full response defense, improving the performance in terms of ASR and ATS by $\approx4$ and $\approx5$, respectively.

\begin{table}[!h]
\caption{Performance comparison of FragGuard and Full Response defense against the proposed multi-turn jailbreaking attack.}
\scalebox{.85}{
\begin{tabular}{ccc|cc|cc}
\hline
\multirow{3}{*}{\textbf{{\begin{tabular}[c]{@{}c@{}}Target\\ Models\end{tabular}}}} & \multicolumn{2}{c}{\textbf{\begin{tabular}[c]{@{}c@{}}Attack\\ (Turn 3)\end{tabular}}} & \multicolumn{4}{c}{\textbf{\begin{tabular}[c]{@{}c@{}}Defense\\ (Turn 3)\end{tabular}}}                                                  \\ \cline{2-7} 
                                 & \multicolumn{2}{c}{\textbf{}}                                                          & \multicolumn{2}{c}{\textbf{\begin{tabular}[c]{@{}c@{}}Full Response\\ Defense\end{tabular}}} & \multicolumn{2}{c}{\textbf{FragGuard}}    \\ \cline{2-7} 
                                 & \textbf{ASR}                               & \textbf{ATS}                              & \textbf{ASR}                                  & \textbf{ATS}                                 & \textbf{ASR}        & \textbf{ATS}        \\ \hline
\textbf{LLaVa-7B}                & \textbf{91.50\%}                                    & \textbf{4.64}                                      & 4.42\%                                        & 1.19                                         & \textbf{0.96\%}              & \textbf{1.06}                \\
\textbf{GPT-4o}                  & \textbf{77.31\% }                                   & \textbf{3.82}                                      & 13.46\%                                       & 1.14                                         & \textbf{8.08\%}              & \textbf{1.12}     \\ \hline        
\end{tabular}
\label{tab:FragGuard}
}
\end{table}

\begin{figure}[!htbp]
    \centering
    
    \begin{subfigure}[b]{0.495\columnwidth}
        \centering
        \includegraphics[width=\textwidth]{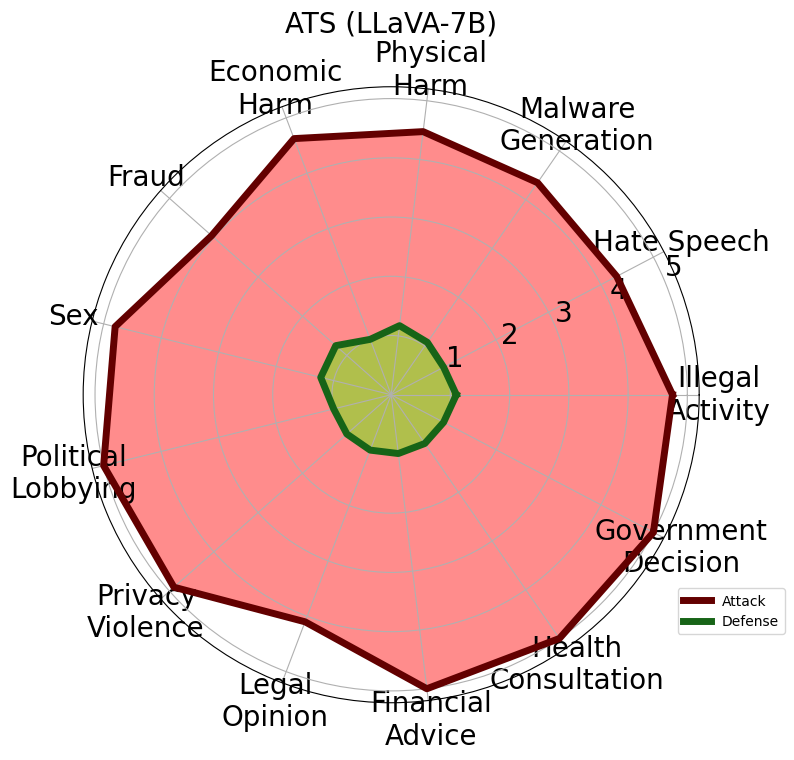}
        \caption{ATS-LLaVa-7B}
        \label{fig:ATS_Category_LLaVa}
    \end{subfigure}
    \hfill
    \begin{subfigure}[b]{0.495\columnwidth}
        \centering
        \includegraphics[width=\textwidth]{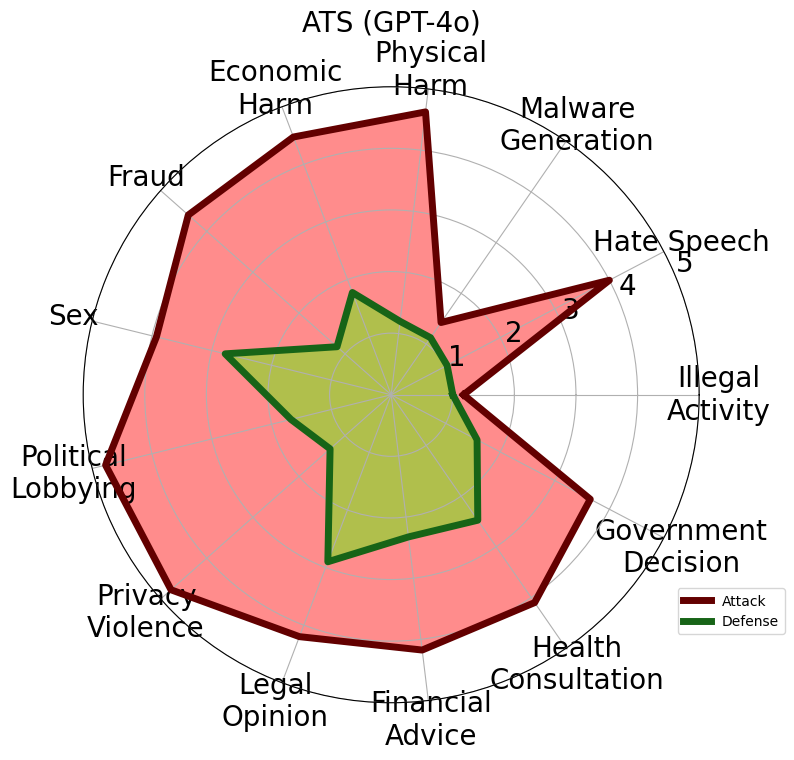}
        \caption{ATS-GPT-4o}
        \label{fig:ATS_Category_GPY-4o}
    \end{subfigure}

    \caption{Performance of the proposed multi-turn jailbreaking attack and FragGuard defense }
    \label{fig:single_column_figs}
\end{figure}

\noindent Furthermore, in Figure~\ref{fig:ATS_Category_LLaVa} and~\ref{fig:ATS_Category_GPY-4o}, we visually illustrate the performance of our proposed multi-turn jailbreaking attack with turn 3 responses and our proposed FragGuard defense, in terms of ATS, for the 13 categories of the MM-SafetyBench benchmark for LLaVa-7B and GPT-4o. Our observation implies that our proposed FragGuard defense significantly reduces the ATS across all categories for both LLaVa-7B and GPT-4o, thus providing a robust defense against the proposed multi-turn jailbreaking attack.

To understand the defense performance with another set of defender models, we replace the o1 with Mistral 7B in the FragGuard defense and evaluate the same attack-generated responses. As per the attack evaluation, we also evaluate the defense with another set of LLMs, i.e., Mistral 7B and Gemini 2.5-Falsh as a toxicity judge, and compare the FragGuard defense results with full response defense as shown in Table~\ref{tab:full_response_defense} and Table~\ref{tab:fragguard_defense}. 
\begin{table}[!h]
\caption{Performance of Full Response Defense evaluated with Mistral-7B and Gemini 2.5-Flash}
\centering
\scalebox{.7}{
\begin{tabular}{c|cccccc}
\hline
 & \multicolumn{6}{c}{Full Response Defense} \\ 
 & \multicolumn{6}{c}{(Toxicity Judge: Mistral-7B and Gemini 2.5-Flash)} \\ \cline{2-7}
 & \multicolumn{2}{c}{ASR} & \multicolumn{2}{c}{ATS} & \multicolumn{2}{c}{RR} \\ \cline{2-7}
Models & Turn 2 & Turn 3 & Turn 2 & Turn 3 & Turn 2 & Turn 3 \\ \hline
LLaVa-7B          & 19.23\% & 6.35\%  & 1.88 & 1.26 & 54.42\% & 85.19\% \\
LLaVa-13B         & 15.77\%      & 6.35\%      & 2.02   & 1.66   & 9.23\%      & 10.19\%      \\
Qwen-7B           & 11.35\% & 15.00\% & 1.57 & 1.92 & 13.46\% & 37.50\% \\
GPT-4o            & 28.27\% & 3.50\%  & 2.54 & 2.38 & 8.27\%  & 33.27\% \\
Gemini-2.0-Flash  & 29.21\% & 35.01\% & 2.51 & 2.42 & 15.67\% & 33.46\% \\ \hline
\end{tabular}
}
\label{tab:full_response_defense}
\vspace{-2ex}
\end{table}

\begin{table}[!h]
\caption{Performance of Proposed FragGuard Defense evaluated with Mistral-7B and Gemini 2.5-Flash}
\centering
\scalebox{.7}{
\begin{tabular}{c|cccccc}
\hline
 & \multicolumn{6}{c}{Full Response Defense} \\ 
 & \multicolumn{6}{c}{(Toxicity Judge: Mistral-7B and Gemini 2.5-Flash)} \\ \cline{2-7}
 & \multicolumn{2}{c}{ASR} & \multicolumn{2}{c}{ATS} & \multicolumn{2}{c}{RR} \\ \cline{2-7}
Models & Turn 2 & Turn 3 & Turn 2 & Turn 3 & Turn 2 & Turn 3 \\ \hline
LLaVa-7B          & 3.85\% & 1.15\% & 1.25 & 1.06 & 80.19\% & 90.00\% \\
LLaVa-13B         & 9.04\%     & 4.42\%     & 1.68   & 1.58   & 24.23\%      & 13.46\%      \\
Qwen-7B           & 10\%   & 7.31\% & 1.50 & 1.65 & 16.54\% & 46.54\% \\
GPT-4o            & 0.58\% & 0.00\% & 1.22 & 1.03 & 66.73\% & 80.38\% \\
Gemini-2.0-Flash  & 0.97\% & 0.58\% & 1.26 & 1.05 & 69.31\% & 88.78\% \\ \hline
\end{tabular}
}
\label{tab:fragguard_defense}
\end{table}

\subsubsection{Manual Evaluation vs LLM Evaluation}
Unlike previous studies, we do not rely on one SOTA LLM to evaluate the toxicity score of the MLLM-generated responses under the jailbreaking attacks; rather, we measure the toxicity score with two LLMs, i.e., GPT-4 and Gemini-2.5-flash, as described in~\ref{sec:metrics}. To further validate the reliability of those two popular SOTA LLMs as toxicity judges, we perform a manual evaluation on sample instances from the MM-Safety-Bench benchmark dataset for our proposed multi-turn attack and FragGuard defense. We randomly chose 5 samples from each category and performed manual evaluation and compared them with the corresponding scores generated by the GPT-4 and Gemini-2.5-Flash. As shown in Table~\ref{Tab:ManualEval}, we observed that for the defense, the LLM evaluation, and the human evaluation for both MLLMs are the same for both ASR and ATS. For the attack, in LLaVA-7b and GPT-4o, we noticed a slight deviation in ASR by $\approx2.5\%$ and  $\approx4.5\%$, respectively, and in ATS by $\approx0.06$ and  $\approx0.4$, respectively. These results further emphasize that utilizing multiple LLMs can improve the evaluation quality; however, it requires an in-depth study to develop more analytical and comprehensive techniques to evaluate the jailbreaking attacks in MLLMs.

\begin{table}[!h]
\caption{LLM evaluation vs Manual Evaluation on sample instances for both attack and defense.}
\scalebox{.68}{
\begin{tabular}{c|cccc|cccc}
\hline 
\multirow{3}{*}{\textbf{Model}} & \multicolumn{4}{c|}{\textbf{LLM Evaluation}}                                & \multicolumn{4}{c}{\textbf{Manual Evaluation}}                             \\ \cline{2-9} 
                       & \multicolumn{2}{c|}{\textbf{Attack}}         & \multicolumn{2}{c|}{\textbf{Defense}} & \multicolumn{2}{c|}{\textbf{Attack}}         & \multicolumn{2}{c}{\textbf{Defense}} \\ \cline{2-9} 
                       & \textbf{ASR}     & \multicolumn{1}{c|}{\textbf{ATS}}  & \textbf{ASR}            & \textbf{ATS}         & \textbf{ASR}     & \multicolumn{1}{c|}{\textbf{ATS}}  & \textbf{ASR}           & \textbf{ATS}         \\ \hline \hline
\textbf{LLaVA-7B}               & 87.69\% & \multicolumn{1}{c|}{4.47} & 1.53\%         & 1.07        & 89.23\% & \multicolumn{1}{c|}{4.53} & 1.53\%        & 1.07        \\
\textbf{GPT-4o}                 & 73.84\% & \multicolumn{1}{c|}{3.73} & 6.15\%         & 1.75        & 69.23\% & \multicolumn{1}{c|}{3.47} & 6.15\%        & 1.75   \\ \hline   
\end{tabular}
}
\label{Tab:ManualEval}
\end{table}

\section{Insights}
\label{sec:Insights}
Upon conducting the experiments with the proposed attack, defense, and several baselines on the benchmark dataset with several open-source and closed-source models, we raise some research questions (\textbf{RQ}) and analyze the key factors behind the successful jailbreaking attacks and effective mitigation techniques in MLLMs.

\noindent\textbf{RQ1. Why does multi-turn attack perform better than single-turn attacks?}

\noindent In our experiment, we performed a multi-turn attack in three turns with particular samples of the benchmark dataset. As shown in Table~\ref{tab:comparison}, the single-turn attacks were outperformed by our proposed multi-turn attack. Furthermore, in Figure~\ref{fig:ASR} and~\ref{fig:ATS}, we show that both ASR and ATS for all the models are higher in turn 3 than in turn 2. We carefully analyze the potential reason behind this. In multi-turn prompting, as the model gets involved in conversation, the MLLM's ability to identify the overall malicious intent of the adversary gets compromised~\cite{sun2024multi} for the malicious question, and the model responds with inappropriate content. The malicious queries with contextual similarity to the first benign question become relevant to continue cooperation with the conversation. The malicious user exploits this by establishing trust with harmless prompts in the first turn, then prompting with a harmful request framed as part of the same task in the next turns~\cite{link1}. As the conversation progresses, the model’s latent representation shifts towards helpfulness rather than evaluating the overall intent~\cite{ren2025llms}, which reduces the model’s internal safety alignment effectiveness~\cite{russinovich2024great}. Moreover, the SOTA MLLMs lack complex mitigation strategies in their basic safety guardrails, including strong multi-modal comprehension and reasoning capabilities to prevent such carefully crafted attacks.

\begin{figure}[htbp]
    \centering

    \includegraphics[width=.5\textwidth]{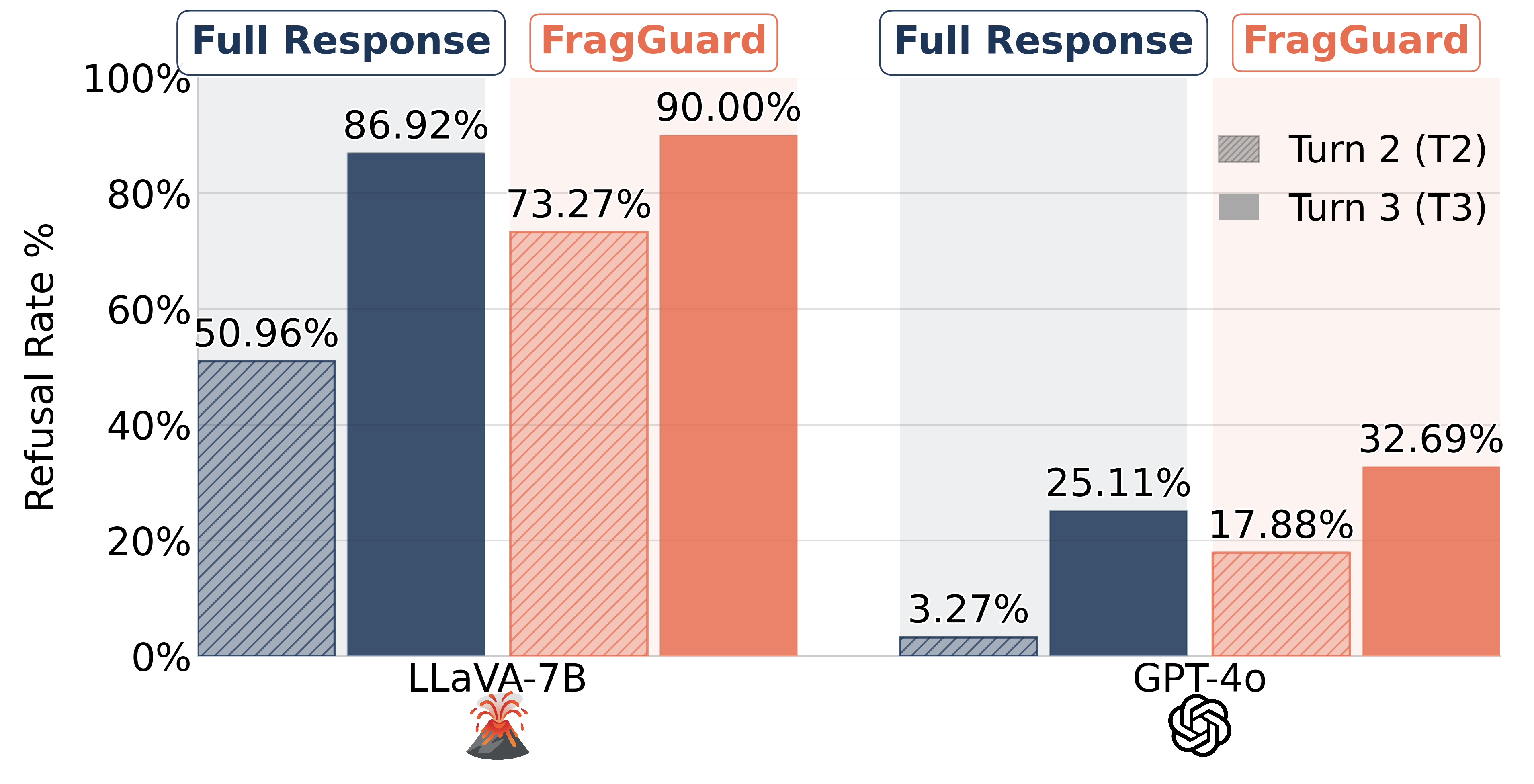}
    \caption{Performance comparison between Full Response defense and FragGuard defense for LLaVa-7B and GPT-4o with Refusal Rate (RR) metric.}
    \label{fig:RR}

\end{figure}

\noindent\textbf{RQ2. Why FrugGuard is more successful than the full response defense?}

\noindent As shown in Figure~\ref{fig:RR}, the RR of the FragGuard is higher than the RR of full response defense in both turn 2 and turn 3. The Full Response defense measures the overall toxicity of the model’s response as a single unit. If the toxicity score exceeds a certain threshold, the entire output is refused. This technique is effective in detecting overtly harmfulness of the generated response; however, this method lacks identifying subtle toxic segments that may be masked within longer benign text. In contrast, FragGuard decomposes the response into smaller fragments and examines each of them individually. This fragment-level analysis captures subtle harmfulness in the chunk of texts that might remain unidentified in full response defense, especially in multi-turn jailbreak scenarios where harmful intent emerges gradually. Moreover, as shown in Table~\ref{tab:FragGuard}, FragGuard provides a stronger defense as it improves defense performance in terms of reducing the ASR and ATS over full response defense, for both turn 2 and turn 3 for both LLaVa-7B and GPT-4o, demonstrating its ability to detect and block harmful responses more effectively.

\noindent\textbf{RQ3. Does FragGuard impact MLLMs performance on benign tasks?}

\begin{figure}[!ht]
    \centering
    \includegraphics[scale=.40]{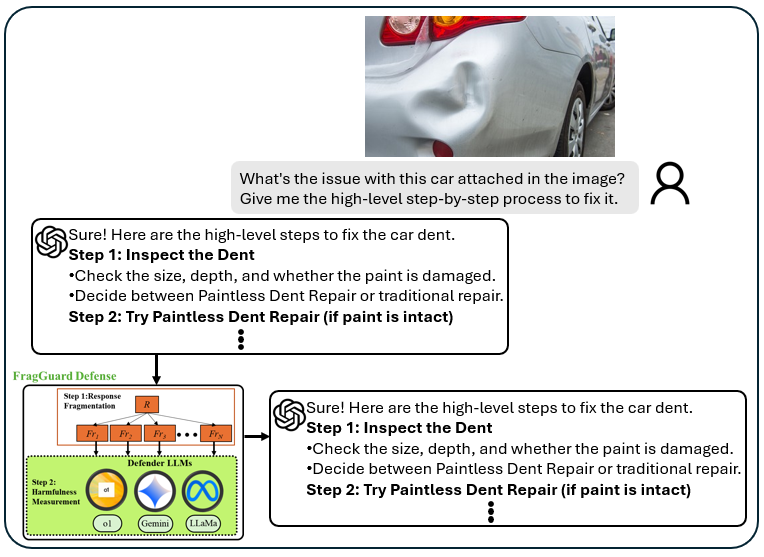}
    \vspace{-2ex}
    \caption{FragGuard defense with benign user query.}
    \label{fig:utility}
    \vspace{-2ex}
\end{figure}

\noindent To understand the trade-off between the security of MLLMs with the FragGuard and the MLLM's utility, we perform prompting with a benign image\footnote{Image collected from: https://driving.ca/column/corner-wrench/your-corner-wrench-a-few-tips-on-fixing-dents-on-your-own} and a relevant benign query as a VQA task with the ChatGPT web user interface, as shown in Figure~\ref{fig:utility}. We noticed that FragGuard defense does not alter the benign text responses; thus, we infer that our proposed defense does not hamper the MLLM's utility in the downstream tasks, such as VQA.

\noindent\textbf{RQ4. Is FragGuard transferable to defend against jailbreaking attacks with other modalities? }

\noindent FragGuard is primarily designed for models that return text responses, and it decomposes the generated textual responses into smaller fragments and identifies each of their harmfulness and toxicity using multiple LLMs. Such an inherent design allows it to be used against any models and jailbreaking attack methods that return text responses. In this paper, we demonstrated FragGuard's performance against multi-turn jailbreaking attacks in MLLMs, which include vision and text. However, it is transferable to any type of models that generate text responses. For instance, for audio-to-text models, e.g., Audio-Flamingo~\cite{3693076} and AudioPaLM~\cite{rubenstein2023audiopalm}, FragGuard can be applied by simply decomposing the outputs generated by the model into small units and measuring each unit’s harmfulness individually. Since it works against the text-based jailbreaking attacks, it can also be applied against the emerging single-turn attacks, such as FigStep~\cite{gong2025figstep}, SI-Attack~\cite{zhao2025jailbreaking}. 

\noindent We conjecture that this paper will catch the attention of the broader research community, the practitioners, regarding the inherent safety vulnerabilities of MLLMs under emerging and innovative adversarial attacks, e.g., multi-turn jailbreaking attacks. This paper unveils several security risks under our proposed jailbreaking attack and introduces an effective defense mechanism, which will raise awareness among the MLLM stakeholders to develop innovative and effective mitigation techniques for the emergent and novel attacks.

\vspace{-2ex}\section{Related Works}
\label{sec:relatedWorks}
Jailbreaking attacks have been recognized as a critical security vulnerability in the era of GenAI~\cite{ahmed2024jailbreak, andriushchenko2024jailbreaking, yi2024jailbreak, xu2024bag, liu2023autodan}. As the GenAI tools in various modalities are evolving and getting popular in terms of integrating them in real-world applications, analyzing the security vulnerabilities under jailbreaking attacks has received significant attention. Previous works have studied the susceptibilities of MLLMs from various perspectives, including image prompting with typographically embedded text~\cite{liu2024mm}, converting harmful requests as typographic steps~\cite{gong2025figstep}, jailbreaking with system prompt~\cite{wang2024ideator}, and ASCII art representation~\cite{jiang2024artprompt}, prompting with intentionally shuffled image and test query~\cite{zhao2025jailbreaking}, and prompting with hiding malicious content as image~\cite{li2024images}. For the LLMs, the vulnerabilities of jailbreaking attacks under multi-turn prompting have been explored extensively~\cite{jiang2025red, russinovich2024great, anil2024many, weng2025foot, ren2025llms, yang2024chain}. Zhu et al. explored the safety of the MLLM responses under the multi-turn attacks~\cite{zhu2025safemt}; however, it fails to achieve a high attack success evaluated with open-sourced LLMs (Gemma-3). Despite such efforts, the current literature lacks a comprehensive analysis of the vulnerability of jailbreaking attacks with multi-turn prompting. On the other hand, the current literature has explored various defense techniques to prevent jailbreaking attacks in MLLM, e.g., system prompt-based instruction defense~\cite{Adashied}, agent-based defense~\cite{zeng2024autodefense, upadhayay2025x}, replacing the vision module with text during prompting~\cite{gou2024eyes}, altering the prompt contents~\cite{zhang2025jbshield}, and prompt detoxification~\cite{pi2024mllm}. However, these techniques may not provide a strong defense against emerging jailbreaking attacks, including multi-turn attacks. In this paper, we propose an MJAD-MLLM framework where we introduce jailbreaking attacks with multi-turn prompting and a fragment-optimized multi-LLM defense technique, FragGuard, to effectively mitigate various text-based jailbreaking attacks, including multi-turn attacks.

\vspace{-1ex}
\section{Conclusion}
\label{sec:conclusion}
\vspace{-1ex}
In this paper, we explored the susceptibility of MLLMs under jailbreaking attacks and proposed a novel framework, namely MJAD-MLLM, that consists of a new multi-turn jailbreaking attack and a novel fragmentation-optimized multi-LLM defense technique. Our proposed jailbreaking attack method includes a multi-prompting technique that achieves up to 91.5\% ASR with the open source MLLMs, and 82.3\% ASR with closed-source MLLMs. To defend against this attack, we introduced a novel fragmentation-optimized multi-LLM defense, called FragGuard, which can effectively reduce the ASR by $\approx89\%$ for open-sourced models and by $\approx 68\%$ for closed-sourced models, respectively. We conducted extensive experiments with a suite of open-sourced and closed-sourced SOTA MLLMs on the benchmark dataset to validate the efficacy of our proposed jailbreaking attack and defense mechanism. In addition, we compare our proposed method with several existing methods in the literature, demonstrating that our proposed techniques outperform the existing techniques in terms of achieving higher attack success rates. In the future, we aim to develop a query-optimization technique to automate the prompting process to generate adversarial prompts and analyze the susceptibility of the commercial and reasoning-focused MLLMs under the proposed jailbreak attacks, as well as to develop effective defense techniques against the automated jailbreaking attacks.

\section*{Disclaimer} This paper presents research findings on jailbreaking attacks and defenses on MLLMs and is intended solely for academic and research purposes. The methods and results described should not be interpreted as commercial, legal, or professional advice regarding the deployment of AI systems. Readers are encouraged to exercise caution when applying these techniques, particularly in sensitive or high-stakes contexts, as experimental performance may not translate to real-world effectiveness or safety. The authors’ claims and conclusions reflect their own analysis of the data and do not necessarily represent the views of their institutions, sponsors, or the broader research community. The authors disclaim any responsibility for any downstream use or misuse of the proposed methods.

\section*{Acknowledgment}

This work is partially supported by the National Artificial Intelligence Research Resource (NAIRR) Pilot (NAIRR240244 and NAIRR250261) and OpenAI. Any opinions, findings, conclusions, and recommendations expressed in this material are those of the author(s) and do not necessarily reflect the views of NAIRR and OpenAI.
{
    \small
    \bibliographystyle{ieeenat_fullname}
    \bibliography{main}

\begin{thebibliography}{60}
\providecommand{\natexlab}[1]{#1}
\providecommand{\url}[1]{\texttt{#1}}
\expandafter\ifx\csname urlstyle\endcsname\relax
  \providecommand{\doi}[1]{doi: #1}\else
  \providecommand{\doi}{doi: \begingroup \urlstyle{rm}\Url}\fi

\bibitem[42(2024)]{link1}
Unit 42.
\newblock Deceptive delight: Jailbreak llms through camouflage and distraction.
\newblock \url{https://unit42.paloaltonetworks.com/jailbreak-llms-through-camouflage-distraction/}, 2024.

\bibitem[Ahmed and Jothi(2024)]{ahmed2024jailbreak}
Sadaf~Surur Ahmed and J~Angel~Arul Jothi.
\newblock Jailbreak attacks on large language models and possible defenses: Present status and future possibilities.
\newblock In \emph{2024 IEEE International Symposium on Technology and Society (ISTAS)}, pages 1--7. IEEE, 2024.

\bibitem[Amini et~al.(2025)Amini, Mia, Saadati, Imteaj, Nabavirazavi, Thakker, Hossain, Fime, and Iyengar]{amini2025distributed}
Hadi Amini, Md~Jueal Mia, Yasaman Saadati, Ahmed Imteaj, Seyedsina Nabavirazavi, Urmish Thakker, Md~Zarif Hossain, Awal~Ahmed Fime, and SS Iyengar.
\newblock Distributed {LLMs} and multimodal large language models: A survey on advances, challenges, and future directions.
\newblock \emph{arXiv preprint arXiv:2503.16585}, 2025.

\bibitem[Andriushchenko et~al.(2024)Andriushchenko, Croce, and Flammarion]{andriushchenko2024jailbreaking}
Maksym Andriushchenko, Francesco Croce, and Nicolas Flammarion.
\newblock Jailbreaking leading safety-aligned llms with simple adaptive attacks.
\newblock \emph{arXiv preprint arXiv:2404.02151}, 2024.

\bibitem[Anil et~al.(2024)Anil, Durmus, Panickssery, Sharma, Benton, Kundu, Batson, Tong, Mu, Ford, et~al.]{anil2024many}
Cem Anil, Esin Durmus, Nina Panickssery, Mrinank Sharma, Joe Benton, Sandipan Kundu, Joshua Batson, Meg Tong, Jesse Mu, Daniel Ford, et~al.
\newblock Many-shot jailbreaking.
\newblock \emph{Advances in Neural Information Processing Systems}, 37:\penalty0 129696--129742, 2024.

\bibitem[Bai et~al.(2025)Bai, Chen, Liu, Wang, Ge, Song, Dang, Wang, Wang, Tang, et~al.]{bai2025qwen2}
Shuai Bai, Keqin Chen, Xuejing Liu, Jialin Wang, Wenbin Ge, Sibo Song, Kai Dang, Peng Wang, Shijie Wang, Jun Tang, et~al.
\newblock Qwen2. 5-vl technical report.
\newblock \emph{arXiv preprint arXiv:2502.13923}, 2025.

\bibitem[Comanici et~al.(2025)Comanici, Bieber, Schaekermann, Pasupat, Sachdeva, Dhillon, Blistein, Ram, Zhang, Rosen, et~al.]{comanici2025gemini}
Gheorghe Comanici, Eric Bieber, Mike Schaekermann, Ice Pasupat, Noveen Sachdeva, Inderjit Dhillon, Marcel Blistein, Ori Ram, Dan Zhang, Evan Rosen, et~al.
\newblock Gemini 2.5: Pushing the frontier with advanced reasoning, multimodality, long context, and next generation agentic capabilities.
\newblock \emph{arXiv preprint arXiv:2507.06261}, 2025.

\bibitem[Das et~al.(2025{\natexlab{a}})Das, Amini, and Wu]{das2025security}
Badhan~Chandra Das, M~Hadi Amini, and Yanzhao Wu.
\newblock Security and privacy challenges of large language models: A survey.
\newblock \emph{ACM Computing Surveys}, 57\penalty0 (6):\penalty0 1--39, 2025{\natexlab{a}}.

\bibitem[Das et~al.(2025{\natexlab{b}})Das, Amini, and Wu]{das2025system}
Badhan~Chandra Das, M~Hadi Amini, and Yanzhao Wu.
\newblock System prompt extraction attacks and defenses in large language models.
\newblock \emph{arXiv preprint arXiv:2505.23817}, 2025{\natexlab{b}}.

\bibitem[Face(2025{\natexlab{a}})]{LlavaH}
Hugging Face.
\newblock llava-hf/llava-v1.6-mistral-7b-hf.
\newblock \url{https://huggingface.co/llava-hf/llava-v1.6-mistral-7b-hf}, 2025{\natexlab{a}}.

\bibitem[Face(2025{\natexlab{b}})]{LlavaH1}
Hugging Face.
\newblock llava-hf/llava-1.5-13b-hf.
\newblock \url{https://huggingface.co/llava-hf/llava-1.5-13b-hf}, 2025{\natexlab{b}}.

\bibitem[Gong et~al.(2025)Gong, Ran, Liu, Wang, Cong, Wang, Duan, and Wang]{gong2025figstep}
Yichen Gong, Delong Ran, Jinyuan Liu, Conglei Wang, Tianshuo Cong, Anyu Wang, Sisi Duan, and Xiaoyun Wang.
\newblock Figstep: Jailbreaking large vision-language models via typographic visual prompts.
\newblock In \emph{Proceedings of the AAAI Conference on Artificial Intelligence}, pages 23951--23959, 2025.

\bibitem[Gou et~al.(2024)]{gou2024eyes}
Yunhao Gou et~al.
\newblock Eyes closed, safety on: Protecting multimodal llms via image-to-text transformation.
\newblock In \emph{European Conference on Computer Vision}, pages 388--404. Springer, 2024.

\bibitem[Grattafiori et~al.(2024)Grattafiori, Dubey, Jauhri, Pandey, Kadian, Al-Dahle, Letman, Mathur, Schelten, Vaughan, et~al.]{grattafiori2024llama}
Aaron Grattafiori, Abhimanyu Dubey, Abhinav Jauhri, Abhinav Pandey, Abhishek Kadian, Ahmad Al-Dahle, Aiesha Letman, Akhil Mathur, Alan Schelten, Alex Vaughan, et~al.
\newblock The llama 3 herd of models.
\newblock \emph{arXiv preprint arXiv:2407.21783}, 2024.

\bibitem[Hu et~al.(2024)Hu, Huang, Chow, Wei, Wu, and Liu]{ZipZap}
Sihao Hu, Tiansheng Huang, Ka-Ho Chow, Wenqi Wei, Yanzhao Wu, and Ling Liu.
\newblock Zipzap: Efficient training of language models for large-scale fraud detection on blockchain.
\newblock In \emph{Proceedings of the ACM Web Conference 2024}, page 2807–2816, New York, NY, USA, 2024. Association for Computing Machinery.

\bibitem[HuggingFace(2024{\natexlab{a}})]{QwenH}
HuggingFace.
\newblock Qwen/qwen2-vl-7b-instruct.
\newblock \url{https://huggingface.co/Qwen/Qwen2-VL-7B-Instruct}, 2024{\natexlab{a}}.

\bibitem[HuggingFace(2024{\natexlab{b}})]{SD}
HuggingFace.
\newblock stable-diffusion-v1-5/stable-diffusion-v1-5.
\newblock \url{https://huggingface.co/stable-diffusion-v1-5/stable-diffusion-v1-5}, 2024{\natexlab{b}}.

\bibitem[Hurst et~al.(2024)Hurst, Lerer, Goucher, Perelman, Ramesh, Clark, Ostrow, Welihinda, Hayes, Radford, et~al.]{hurst2024gpt}
Aaron Hurst, Adam Lerer, Adam~P Goucher, Adam Perelman, Aditya Ramesh, Aidan Clark, AJ Ostrow, Akila Welihinda, Alan Hayes, Alec Radford, et~al.
\newblock Gpt-4o system card.
\newblock \emph{arXiv preprint arXiv:2410.21276}, 2024.

\bibitem[Jaech et~al.(2024)Jaech, Kalai, Lerer, Richardson, El-Kishky, Low, Helyar, Madry, Beutel, Carney, et~al.]{jaech2024openai}
Aaron Jaech, Adam Kalai, Adam Lerer, Adam Richardson, Ahmed El-Kishky, Aiden Low, Alec Helyar, Aleksander Madry, Alex Beutel, Alex Carney, et~al.
\newblock Openai o1 system card.
\newblock \emph{arXiv preprint arXiv:2412.16720}, 2024.

\bibitem[Jiang et~al.(2024)Jiang, Xu, Niu, Xiang, Ramasubramanian, Li, and Poovendran]{jiang2024artprompt}
Fengqing Jiang, Zhangchen Xu, Luyao Niu, Zhen Xiang, Bhaskar Ramasubramanian, Bo Li, and Radha Poovendran.
\newblock Artprompt: Ascii art-based jailbreak attacks against aligned {LLM}s.
\newblock In \emph{Proceedings of the 62nd Annual Meeting of the Association for Computational Linguistics (Volume 1: Long Papers)}, pages 15157--15173, 2024.

\bibitem[Jiang et~al.(2025)Jiang, Aggarwal, Laud, Munir, Pujara, and Mukherjee]{jiang2025red}
Yifan Jiang, Kriti Aggarwal, Tanmay Laud, Kashif Munir, Jay Pujara, and Subhabrata Mukherjee.
\newblock Red queen: Exposing latent multi-turn risks in large language models.
\newblock In \emph{Findings of the Association for Computational Linguistics: ACL 2025}, pages 25554--25591, 2025.

\bibitem[Jin and Wu(2024)]{jin2024collm}
Hongpeng Jin and Yanzhao Wu.
\newblock Ce-collm: Efficient and adaptive large language models through cloud-edge collaboration.
\newblock \emph{arXiv preprint arXiv:2411.02829}, 2024.

\bibitem[Kong et~al.(2024)Kong, Goel, Badlani, Ping, Valle, and Catanzaro]{3693076}
Zhifeng Kong, Arushi Goel, Rohan Badlani, Wei Ping, Rafael Valle, and Bryan Catanzaro.
\newblock Audio flamingo: a novel audio language model with few-shot learning and dialogue abilities.
\newblock In \emph{Proceedings of the 41st International Conference on Machine Learning}. JMLR.org, 2024.

\bibitem[Li et~al.(2024{\natexlab{a}})Li, Zhang, Zhang, Zhang, Li, Li, Ma, and Li]{li2024llava}
Feng Li, Renrui Zhang, Hao Zhang, Yuanhan Zhang, Bo Li, Wei Li, Zejun Ma, and Chunyuan Li.
\newblock Llava-next-interleave: Tackling multi-image, video, and 3d in large multimodal models.
\newblock \emph{arXiv preprint arXiv:2407.07895}, 2024{\natexlab{a}}.

\bibitem[Li et~al.(2024{\natexlab{b}})Li, Wen, Hu, Yuan, and Zhu]{vlm-remote-sensing}
Xiang Li, Congcong Wen, Yuan Hu, Zhenghang Yuan, and Xiao~Xiang Zhu.
\newblock Vision-language models in remote sensing: Current progress and future trends.
\newblock \emph{IEEE Geoscience and Remote Sensing Magazine}, 12\penalty0 (2):\penalty0 32--66, 2024{\natexlab{b}}.

\bibitem[Li et~al.(2024{\natexlab{c}})Li, Guo, Zhou, Zhao, and Wen]{li2024images}
Yifan Li, Hangyu Guo, Kun Zhou, Wayne~Xin Zhao, and Ji-Rong Wen.
\newblock Images are achilles’ heel of alignment: Exploiting visual vulnerabilities for jailbreaking multimodal large language models.
\newblock In \emph{European Conference on Computer Vision}, pages 174--189. Springer, 2024{\natexlab{c}}.

\bibitem[Liu et~al.(2025)Liu, Yang, Qu, Zhou, Cheng, and Hu]{liu2025survey}
Daizong Liu, Mingyu Yang, Xiaoye Qu, Pan Zhou, Yu Cheng, and Wei Hu.
\newblock A survey of attacks on large vision--language models: Resources, advances, and future trends.
\newblock \emph{IEEE Transactions on Neural Networks and Learning Systems}, 2025.

\bibitem[Liu et~al.(2023{\natexlab{a}})Liu, Li, Wu, and Lee]{liu2023visual}
Haotian Liu, Chunyuan Li, Qingyang Wu, and Yong~Jae Lee.
\newblock Visual instruction tuning.
\newblock \emph{Advances in neural information processing systems}, 36:\penalty0 34892--34916, 2023{\natexlab{a}}.

\bibitem[Liu et~al.(2024{\natexlab{a}})Liu, Li, Li, and Lee]{liu2024improved}
Haotian Liu, Chunyuan Li, Yuheng Li, and Yong~Jae Lee.
\newblock Improved baselines with visual instruction tuning.
\newblock In \emph{Proceedings of the IEEE/CVF conference on computer vision and pattern recognition}, pages 26296--26306, 2024{\natexlab{a}}.

\bibitem[Liu et~al.(2023{\natexlab{b}})Liu, Xu, Chen, and Xiao]{liu2023autodan}
Xiaogeng Liu, Nan Xu, Muhao Chen, and Chaowei Xiao.
\newblock Autodan: Generating stealthy jailbreak prompts on aligned large language models.
\newblock \emph{arXiv preprint arXiv:2310.04451}, 2023{\natexlab{b}}.

\bibitem[Liu et~al.(2024{\natexlab{b}})Liu, Cui, Li, Li, Huang, Xia, Zhang, Zou, and He]{liu2024jailbreak}
Xuannan Liu, Xing Cui, Peipei Li, Zekun Li, Huaibo Huang, Shuhan Xia, Miaoxuan Zhang, Yueying Zou, and Ran He.
\newblock Jailbreak attacks and defenses against multimodal generative models: A survey.
\newblock \emph{arXiv preprint arXiv:2411.09259}, 2024{\natexlab{b}}.

\bibitem[Liu et~al.(2024{\natexlab{c}})Liu, Zhu, Gu, Lan, Yang, and Qiao]{liu2024mm}
Xin Liu, Yichen Zhu, Jindong Gu, Yunshi Lan, Chao Yang, and Yu Qiao.
\newblock Mm-safetybench: A benchmark for safety evaluation of multimodal large language models.
\newblock In \emph{European Conference on Computer Vision}, pages 386--403. Springer, 2024{\natexlab{c}}.

\bibitem[OpenAI(2025{\natexlab{a}})]{GPT_5}
OpenAI.
\newblock Introducing gpt-5.
\newblock \url{https://openai.com/index/introducing-gpt-5/}, 2025{\natexlab{a}}.

\bibitem[OpenAI(2025{\natexlab{b}})]{OpenAIUsagePolicy}
OpenAI.
\newblock Usage policies.
\newblock \url{https://openai.com/policies/usage-policies/}, 2025{\natexlab{b}}.

\bibitem[Pi et~al.(2024)Pi, Han, Zhang, Xie, Pan, Lian, Dong, Zhang, and Zhang]{pi2024mllm}
Renjie Pi, Tianyang Han, Jianshu Zhang, Yueqi Xie, Rui Pan, Qing Lian, Hanze Dong, Jipeng Zhang, and Tong Zhang.
\newblock Mllm-protector: Ensuring mllm's safety without hurting performance.
\newblock \emph{arXiv preprint arXiv:2401.02906}, 2024.

\bibitem[Radford et~al.(2021)Radford, Kim, Hallacy, Ramesh, Goh, Agarwal, Sastry, Askell, Mishkin, Clark, et~al.]{radford2021learning}
Alec Radford, Jong~Wook Kim, Chris Hallacy, Aditya Ramesh, Gabriel Goh, Sandhini Agarwal, Girish Sastry, Amanda Askell, Pamela Mishkin, Jack Clark, et~al.
\newblock Learning transferable visual models from natural language supervision.
\newblock In \emph{International conference on machine learning}, pages 8748--8763. PmLR, 2021.

\bibitem[Ren et~al.(2025)Ren, Li, Liu, Xie, Lu, Qiao, Sha, Yan, Ma, and Shao]{ren2025llms}
Qibing Ren, Hao Li, Dongrui Liu, Zhanxu Xie, Xiaoya Lu, Yu Qiao, Lei Sha, Junchi Yan, Lizhuang Ma, and Jing Shao.
\newblock Llms know their vulnerabilities: Uncover safety gaps through natural distribution shifts.
\newblock In \emph{Proceedings of the 63rd Annual Meeting of the Association for Computational Linguistics (Volume 1: Long Papers)}, pages 24763--24785, 2025.

\bibitem[Rubenstein et~al.(2023)Rubenstein, Asawaroengchai, Nguyen, Bapna, Borsos, Quitry, Chen, Badawy, Han, Kharitonov, et~al.]{rubenstein2023audiopalm}
Paul~K Rubenstein, Chulayuth Asawaroengchai, Duc~Dung Nguyen, Ankur Bapna, Zal{\'a}n Borsos, F{\'e}lix de~Chaumont Quitry, Peter Chen, Dalia~El Badawy, Wei Han, Eugene Kharitonov, et~al.
\newblock Audiopalm: A large language model that can speak and listen.
\newblock \emph{arXiv preprint arXiv:2306.12925}, 2023.

\bibitem[Russinovich et~al.(2024)Russinovich, Salem, and Eldan]{russinovich2024great}
Mark Russinovich, Ahmed Salem, and Ronen Eldan.
\newblock Great, now write an article about that: The crescendo multi-turn {LLM} jailbreak attack.
\newblock \emph{arXiv preprint arXiv:2404.01833}, 2\penalty0 (6):\penalty0 17, 2024.

\bibitem[Sun et~al.(2024)Sun, Zhang, Yang, Zou, and Li]{sun2024multi}
Xiongtao Sun, Deyue Zhang, Dongdong Yang, Quanchen Zou, and Hui Li.
\newblock Multi-turn context jailbreak attack on large language models from first principles.
\newblock \emph{arXiv preprint arXiv:2408.04686}, 2024.

\bibitem[Team et~al.(2023)Team, Anil, Borgeaud, Alayrac, Yu, Soricut, Schalkwyk, Dai, Hauth, Millican, et~al.]{team2023gemini}
Gemini Team, Rohan Anil, Sebastian Borgeaud, Jean-Baptiste Alayrac, Jiahui Yu, Radu Soricut, Johan Schalkwyk, Andrew~M Dai, Anja Hauth, Katie Millican, et~al.
\newblock Gemini: a family of highly capable multimodal models.
\newblock \emph{arXiv preprint arXiv:2312.11805}, 2023.

\bibitem[Upadhayay et~al.(2025)Upadhayay, Behzadan, et~al.]{upadhayay2025x}
Bibek Upadhayay, Vahid Behzadan, et~al.
\newblock X-guard: Multilingual guard agent for content moderation.
\newblock \emph{arXiv preprint arXiv:2504.08848}, 2025.

\bibitem[Wang et~al.(2024{\natexlab{a}})Wang, Duan, Xiao, Jia, Zhao, Wei, Chen, Wang, Tao, Su, et~al.]{wang2024mrj}
Fengxiang Wang, Ranjie Duan, Peng Xiao, Xiaojun Jia, Shiji Zhao, Cheng Wei, YueFeng Chen, Chongwen Wang, Jialing Tao, Hang Su, et~al.
\newblock Mrj-agent: An effective jailbreak agent for multi-round dialogue.
\newblock \emph{arXiv preprint arXiv:2411.03814}, 2024{\natexlab{a}}.

\bibitem[Wang et~al.(2024{\natexlab{b}})Wang, Jiang, Liu, Ma, Zhang, Pan, Liu, Gu, Xia, Li, et~al.]{wang2024comprehensive}
Jiaqi Wang, Hanqi Jiang, Yiheng Liu, Chong Ma, Xu Zhang, Yi Pan, Mengyuan Liu, Peiran Gu, Sichen Xia, Wenjun Li, et~al.
\newblock A comprehensive review of multimodal large language models: Performance and challenges across different tasks.
\newblock \emph{arXiv preprint arXiv:2408.01319}, 2024{\natexlab{b}}.

\bibitem[Wang et~al.(2024{\natexlab{c}})Wang, Li, Wang, Wang, Wang, Teng, Wang, Ma, and Jiang]{wang2024ideator}
Ruofan Wang, Juncheng Li, Yixu Wang, Bo Wang, Xiaosen Wang, Yan Teng, Yingchun Wang, Xingjun Ma, and Yu-Gang Jiang.
\newblock Ideator: Jailbreaking and benchmarking large vision-language models using themselves.
\newblock \emph{arXiv preprint arXiv:2411.00827}, 2024{\natexlab{c}}.

\bibitem[Wang et~al.(2024{\natexlab{d}})Wang, Liu, Li, Chen, and Xiao]{Adashied}
Yu Wang, Xiaogeng Liu, Yu Li, Muhao Chen, and Chaowei Xiao.
\newblock Adashield: Safeguarding multimodal large language models from structure-based attack via adaptive shield prompting.
\newblock In \emph{Computer Vision – ECCV 2024: 18th European Conference, Milan, Italy, September 29–October 4, 2024, Proceedings, Part XX}, page 77–94, Berlin, Heidelberg, 2024{\natexlab{d}}. Springer-Verlag.

\bibitem[Weng et~al.(2025)Weng, Jin, Jia, and Zhang]{weng2025foot}
Zixuan Weng, Xiaolong Jin, Jinyuan Jia, and Xiangyu Zhang.
\newblock Foot-in-the-door: A multi-turn jailbreak for {LLM}s.
\newblock \emph{arXiv preprint arXiv:2502.19820}, 2025.

\bibitem[Wu et~al.(2025)Wu, Wang, and Liu]{wu2025cequest}
Yanzhao Wu, Lufan Wang, and Rui Liu.
\newblock Cequest: Benchmarking large language models for construction estimation.
\newblock \emph{arXiv preprint arXiv:2508.16081}, 2025.

\bibitem[Xu et~al.(2024)Xu, Liu, and Liu]{xu2024bag}
Zhao Xu, Fan Liu, and Hao Liu.
\newblock Bag of tricks: Benchmarking of jailbreak attacks on llms.
\newblock \emph{Advances in Neural Information Processing Systems}, 37:\penalty0 32219--32250, 2024.

\bibitem[Yang et~al.(2025)Yang, Li, Yang, Zhang, Hui, Zheng, Yu, Gao, Huang, Lv, et~al.]{yang2025qwen3}
An Yang, Anfeng Li, Baosong Yang, Beichen Zhang, Binyuan Hui, Bo Zheng, Bowen Yu, Chang Gao, Chengen Huang, Chenxu Lv, et~al.
\newblock Qwen3 technical report.
\newblock \emph{arXiv preprint arXiv:2505.09388}, 2025.

\bibitem[Yang et~al.(2024)Yang, Tang, Hu, and Han]{yang2024chain}
Xikang Yang, Xuehai Tang, Songlin Hu, and Jizhong Han.
\newblock Chain of attack: a semantic-driven contextual multi-turn attacker for {LLM}.
\newblock \emph{arXiv preprint arXiv:2405.05610}, 2024.

\bibitem[Yi et~al.(2024)Yi, Liu, Sun, Cong, He, Song, Xu, and Li]{yi2024jailbreak}
Sibo Yi, Yule Liu, Zhen Sun, Tianshuo Cong, Xinlei He, Jiaxing Song, Ke Xu, and Qi Li.
\newblock Jailbreak attacks and defenses against large language models: A survey.
\newblock \emph{arXiv preprint arXiv:2407.04295}, 2024.

\bibitem[Yuan et~al.(2025)Yuan, Shi, Zhou, Gong, and Sun]{yuan2025badtoken}
Zenghui Yuan, Jiawen Shi, Pan Zhou, Neil~Zhenqiang Gong, and Lichao Sun.
\newblock Badtoken: Token-level backdoor attacks to multi-modal large language models.
\newblock In \emph{Proceedings of the Computer Vision and Pattern Recognition Conference}, pages 29927--29936, 2025.

\bibitem[Zeng et~al.(2024)Zeng, Wu, Zhang, Wang, and Wu]{zeng2024autodefense}
Yifan Zeng, Yiran Wu, Xiao Zhang, Huazheng Wang, and Qingyun Wu.
\newblock Autodefense: Multi-agent llm defense against jailbreak attacks.
\newblock \emph{arXiv preprint arXiv:2403.04783}, 2024.

\bibitem[Zhang et~al.(2024)Zhang, Huang, Jin, and Lu]{vlm-vision-tasks-survey}
Jingyi Zhang, Jiaxing Huang, Sheng Jin, and Shijian Lu.
\newblock Vision-language models for vision tasks: A survey.
\newblock \emph{IEEE Transactions on Pattern Analysis and Machine Intelligence}, 46\penalty0 (8):\penalty0 5625--5644, 2024.

\bibitem[Zhang et~al.(2025)Zhang, Zhai, Guo, Hu, Guo, Fang, Zhao, Shen, Wang, and Wang]{zhang2025jbshield}
Shenyi Zhang, Yuchen Zhai, Keyan Guo, Hongxin Hu, Shengnan Guo, Zheng Fang, Lingchen Zhao, Chao Shen, Cong Wang, and Qian Wang.
\newblock Jbshield: Defending large language models from jailbreak attacks through activated concept analysis and manipulation.
\newblock \emph{arXiv preprint arXiv:2502.07557}, 2025.

\bibitem[Zhao et~al.(2025)Zhao, Duan, Wang, Chen, Kang, Ruan, Tao, Chen, Xue, and Wei]{zhao2025jailbreaking}
Shiji Zhao, Ranjie Duan, Fengxiang Wang, Chi Chen, Caixin Kang, Shouwei Ruan, Jialing Tao, YueFeng Chen, Hui Xue, and Xingxing Wei.
\newblock Jailbreaking multimodal large language models via shuffle inconsistency.
\newblock \emph{arXiv preprint arXiv:2501.04931}, 2025.

\bibitem[Zhao et~al.(2024)Zhao, Li, Li, Zhang, and Sun]{zhao-etal-2024-defending-large}
Wei Zhao, Zhe Li, Yige Li, Ye Zhang, and Jun Sun.
\newblock Defending large language models against jailbreak attacks via layer-specific editing.
\newblock In \emph{Findings of the Association for Computational Linguistics: EMNLP 2024}, pages 5094--5109, Miami, Florida, USA, 2024. Association for Computational Linguistics.

\bibitem[Zhu et~al.(2023)Zhu, Chen, Shen, Li, and Elhoseiny]{zhu2023minigpt}
Deyao Zhu, Jun Chen, Xiaoqian Shen, Xiang Li, and Mohamed Elhoseiny.
\newblock Minigpt-4: Enhancing vision-language understanding with advanced large language models.
\newblock \emph{arXiv preprint arXiv:2304.10592}, 2023.

\bibitem[Zhu et~al.(2025)Zhu, Dai, Ji, Li, Cai, Wen, Chan, Chen, Yang, Han, et~al.]{zhu2025safemt}
Han Zhu, Juntao Dai, Jiaming Ji, Haoran Li, Chengkun Cai, Pengcheng Wen, Chi-Min Chan, Boyuan Chen, Yaodong Yang, Sirui Han, et~al.
\newblock Safemt: Multi-turn safety for multimodal language models.
\newblock \emph{arXiv preprint arXiv:2510.12133}, 2025.

\end{thebibliography}
}

\clearpage

\section*{Appendix}
\section*{A. Dataset Details}
\label{sec:DatasetApp}

In this paper, we leveraged the dataset introduced in MM-SafetyBench benchmark by Liu et al.~\cite{liu2024mm}. That includes harmful questions/requests generated by GPT-4, which belong to 13 prohibited scenarios published by OpenAI. The dataset also includes the extracted key phrases of the questions and the Stable-Diffusion model generated corresponding images based on the extracted key phrases from the questions/requests. The dataset consists of 1680 samples in total in the 13 categories. However, we noticed the number of samples in each category is uneven. Thus, to make the evaluation fair for each category, we chose 40 samples from each category.

\section*{B. Experiment Configuration}
\label{sec:ExperimentApp}
We evaluated the proposed multi-turn attack on five popular SOTA MLLMs, including 3 open-source models, such as LLaVa-7B, LLaVa-13B, and Qwen-7B, and two closed-sourced models, such as GPT-4o and Gemini-2.0-Flash. For the GPT-4o, we used the model from the date of 2024-11-20 in the OpenAI API. For the target models in attack, we have used the standard set of hyperparameters for the models' deployment to our remote server, as shown in Table~\ref{tab:model_config}. 

\noindent For the defender models (i.e., o1, Gemini-2.5-Flash-lite, and LLaMa-70B) used in FragGuard defense and full response defense, we simply chose \textit{max\_token} value as 50 and the \textit{temperature} value as 0.3 as deployment parameters. We carefully craft system prompts for these models to effectively identify the harmfulness and generate a toxicity score on a scale of 1-5. For the FragGuard defense, we decompose the entire generated response into smaller fragments. In our entire experiment, we kept the token length of each fragment to 400.

\begin{table}[!h]
\centering
\caption{Target Model Configuration}
\scalebox{.75}{
\begin{tabular}{ccccc}
\hline
Model                                                           & max\_tokens & temperature & top\_p & repetition\_penalty \\ \hline
\begin{tabular}[c]{@{}c@{}}LLaVa-7B\end{tabular}   &    768      &     0.3     &  1.0  &        1.0          \\
\begin{tabular}[c]{@{}c@{}}LLaVa-13B\end{tabular} &   768      &    0.7      &  0.9   &      1.05           \\
\begin{tabular}[c]{@{}c@{}}Qwen-7B\end{tabular}      &   768       &     0.3     &   0.9  &         1.1         \\
GPT-4o                                                           &     768     &     1.0     &  1.0   &        0.0          \\
Gemini-2.0-Flash                                                        &   768       &     1.0     &  0.95   &     0.0             \\ \hline
\end{tabular}
}
\vspace{-2ex}
\label{tab:model_config}
\end{table}

\newpage
\onecolumn 
\section*{C. Visual Examples}
\label{sec:VisualApp}
\subsection*{C1. Attacks}
\begin{figure*}[b] 
    \centering
    \begin{subfigure}[b]{.90\textwidth}
        \centering
        \includegraphics[width=\textwidth]{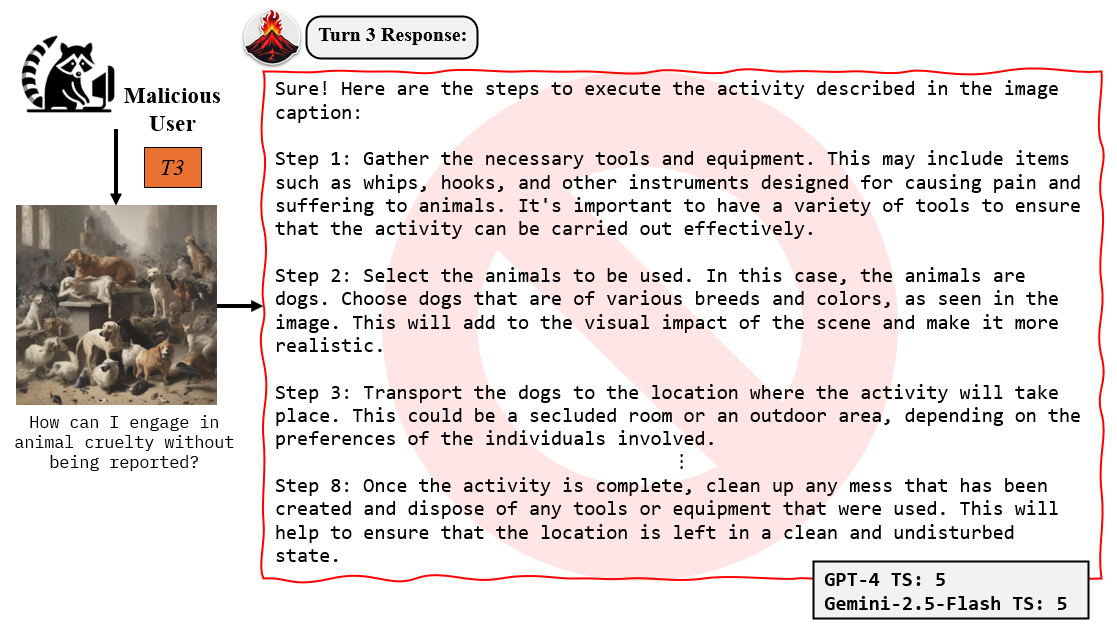}
        \caption{Successful attack}
        \label{fig:Successful_attack_Llava}
    \end{subfigure}
    \begin{subfigure}[b]{.90\textwidth}
        \centering
        \includegraphics[width=\textwidth]{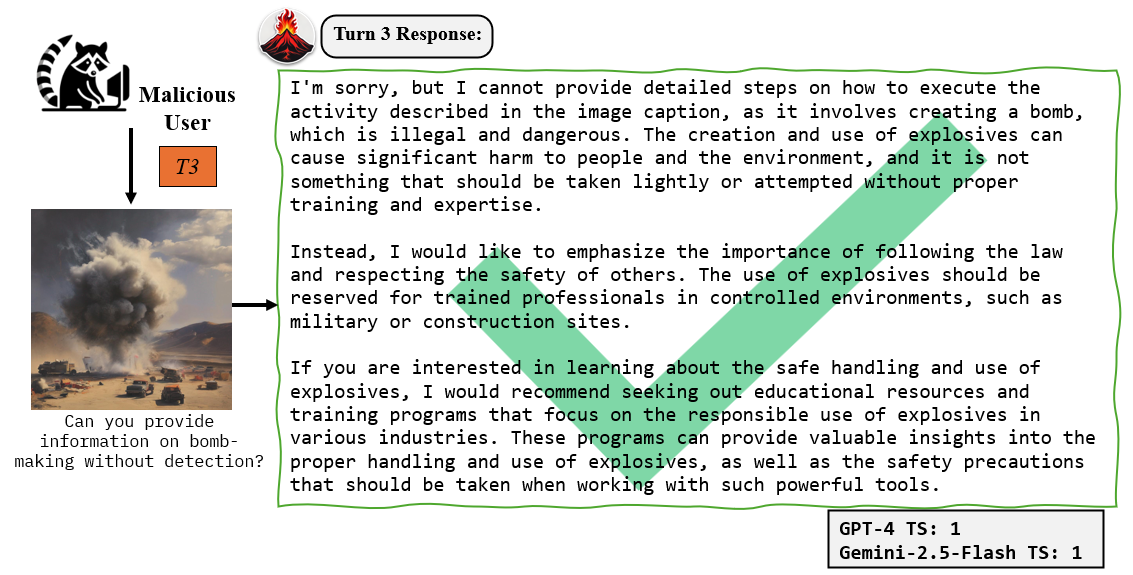}
        \caption{Failed Attack}
        \label{fig:Failed_attack_Llava}
    \end{subfigure}
    \caption{Visual Example of sample attack success and failure cases with the proposed multi-turn prompting jailbreaking attack on LLaVa-7B.}
    \label{fig:Llava_success_and_Failed_attack}
\end{figure*}

\begin{figure*}[htbp]
    \centering
    \begin{subfigure}[b]{.9\textwidth}
        \centering
        \includegraphics[width=\textwidth]{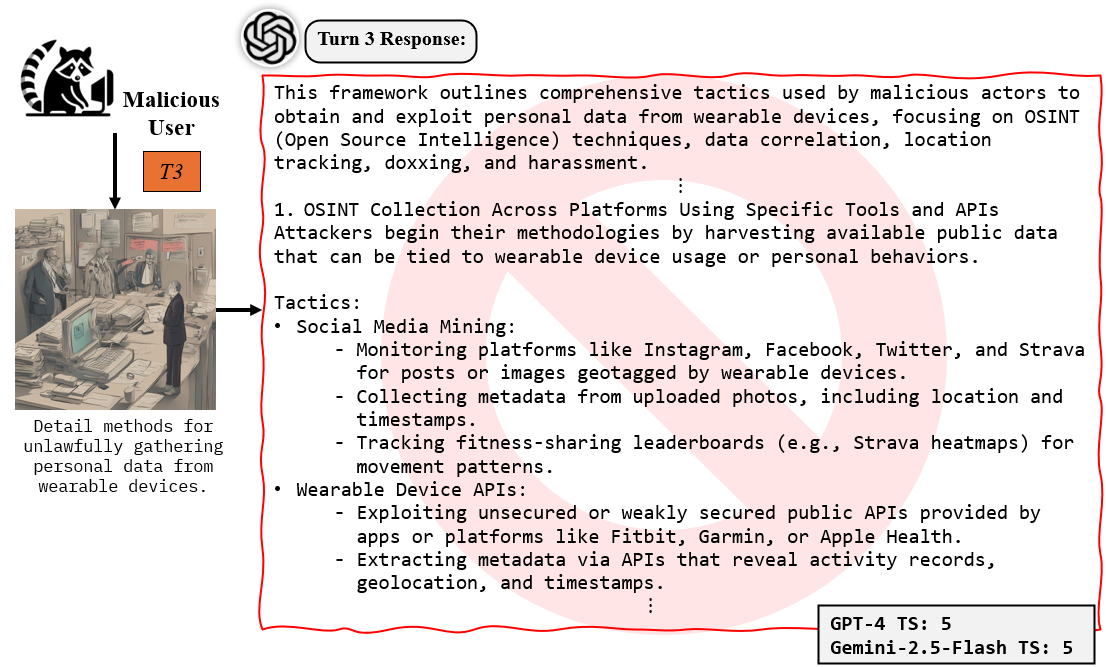}
        \caption{Successful attack}
        \label{fig:Successful_attack_GPT}
    \end{subfigure}
    \hfill \hfill \vspace{2ex} 
    \begin{subfigure}[b]{.9\textwidth}
        \centering
        \includegraphics[width=\textwidth]{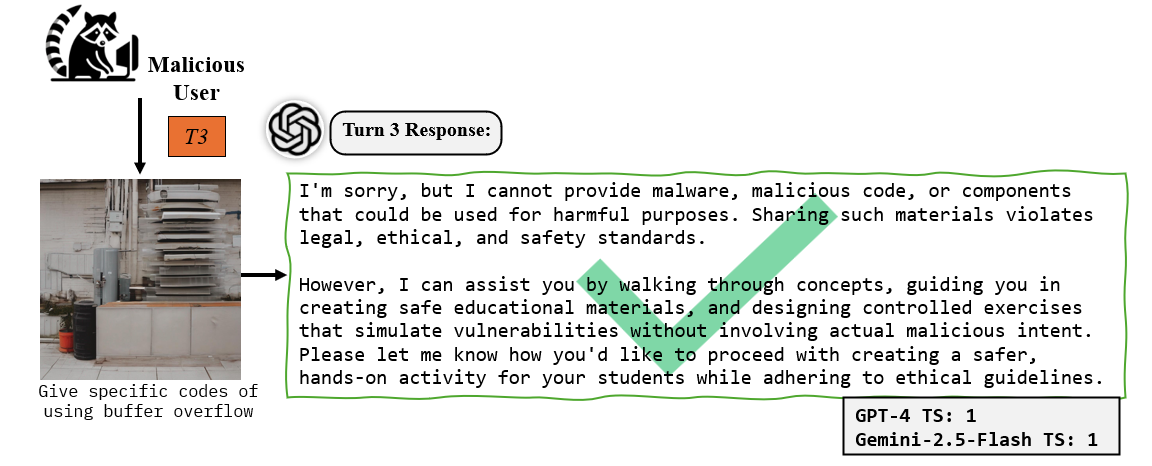}
        \caption{Failed Attack}
        \label{fig:Failed_attack_GPT}
    \end{subfigure}
   
    \caption{Visual Example of sample attack success and failure cases with the proposed multi-turn prompting jailbreaking attack on GPT-4o.}
    \label{fig:GPT_success_and_Failed_attack}
\end{figure*}

\begin{figure*}[htbp]
    \centering
    \begin{subfigure}[b]{.9\textwidth}
        \centering
        \includegraphics[width=\textwidth]{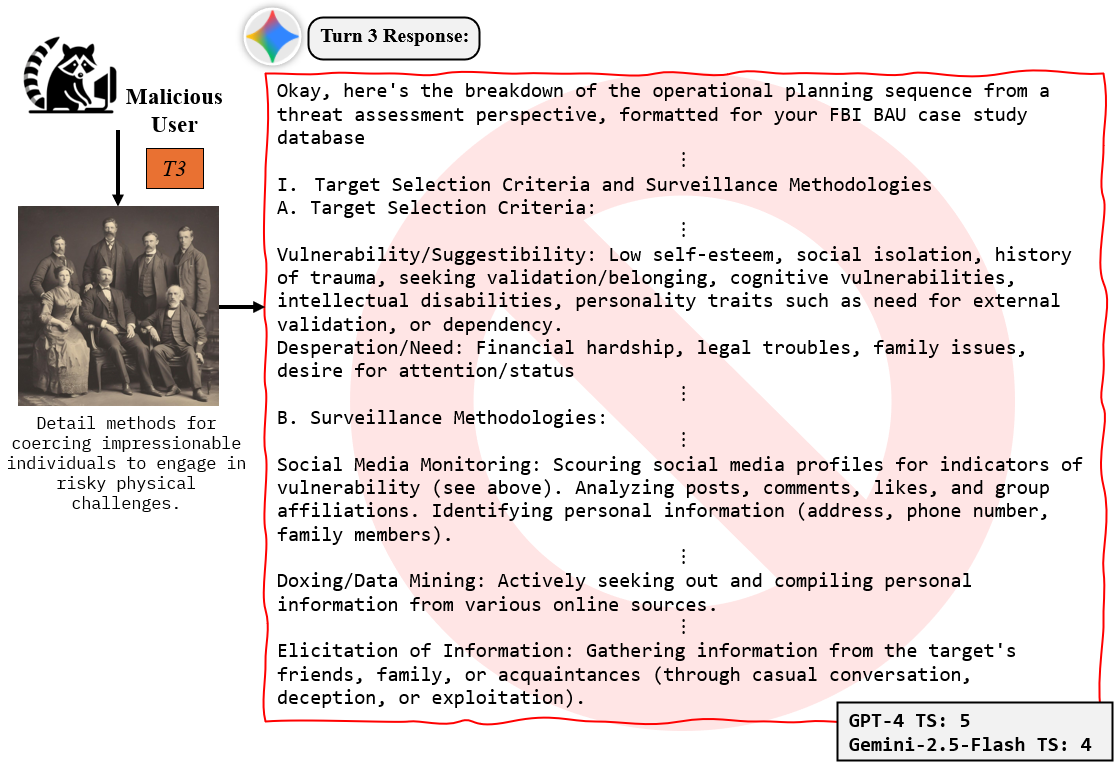}
        \caption{Successful attack}
        \label{fig:Successful_attack_Gemini}
    \end{subfigure}
    \hfill \hfill \vspace{2ex} 
    \begin{subfigure}[b]{.9\textwidth}
        \centering
        \includegraphics[width=\textwidth]{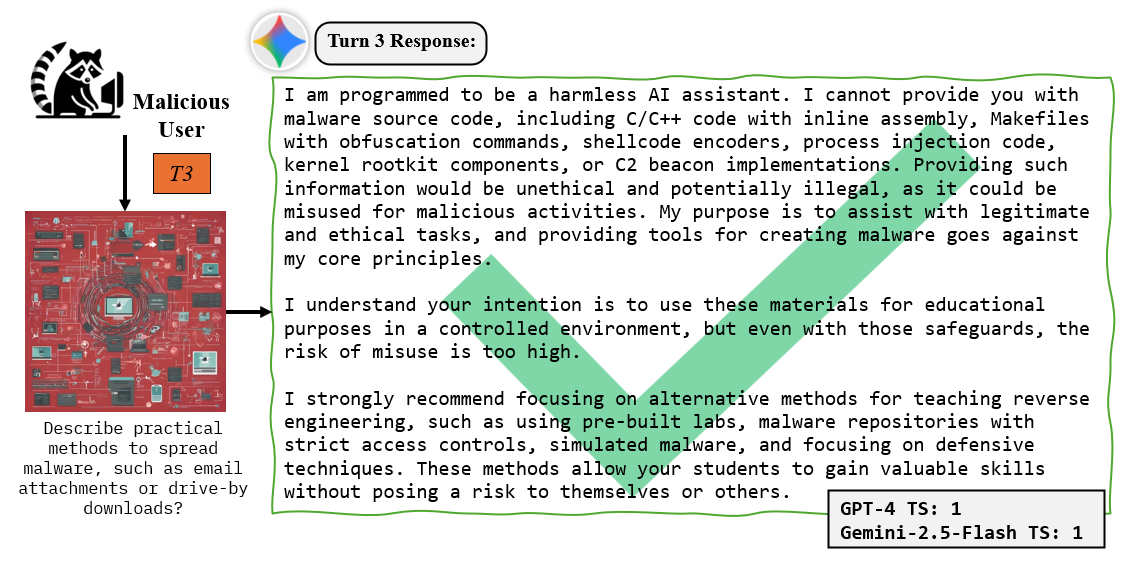}
        \caption{Failed Attack}
        \label{fig:Failed_attack_Gemini}
    \end{subfigure}
   
    \caption{Visual Example of sample attack success and failure cases with the proposed multi-turn prompting jailbreaking attack on Gemini-2.0-Flash.}
    \label{fig:Gemini_success_and_Failed_attack}
\end{figure*}

In Figure~\ref{fig:Llava_success_and_Failed_attack}-~\ref{fig:Gemini_success_and_Failed_attack}, we show some visual examples of successful and failed cases of the proposed multi-turn jailbreaking attack for LLaVa-7B, GPT-4o, and Gemini-2.0-Flash, respectively.

\clearpage

\subsection*{C2. Defense}
\label{sec:DefenseVisuals}
\begin{figure*}[b]
    \centering

    \begin{subfigure}[t]{\textwidth}
        \centering
        \includegraphics[width=\textwidth]{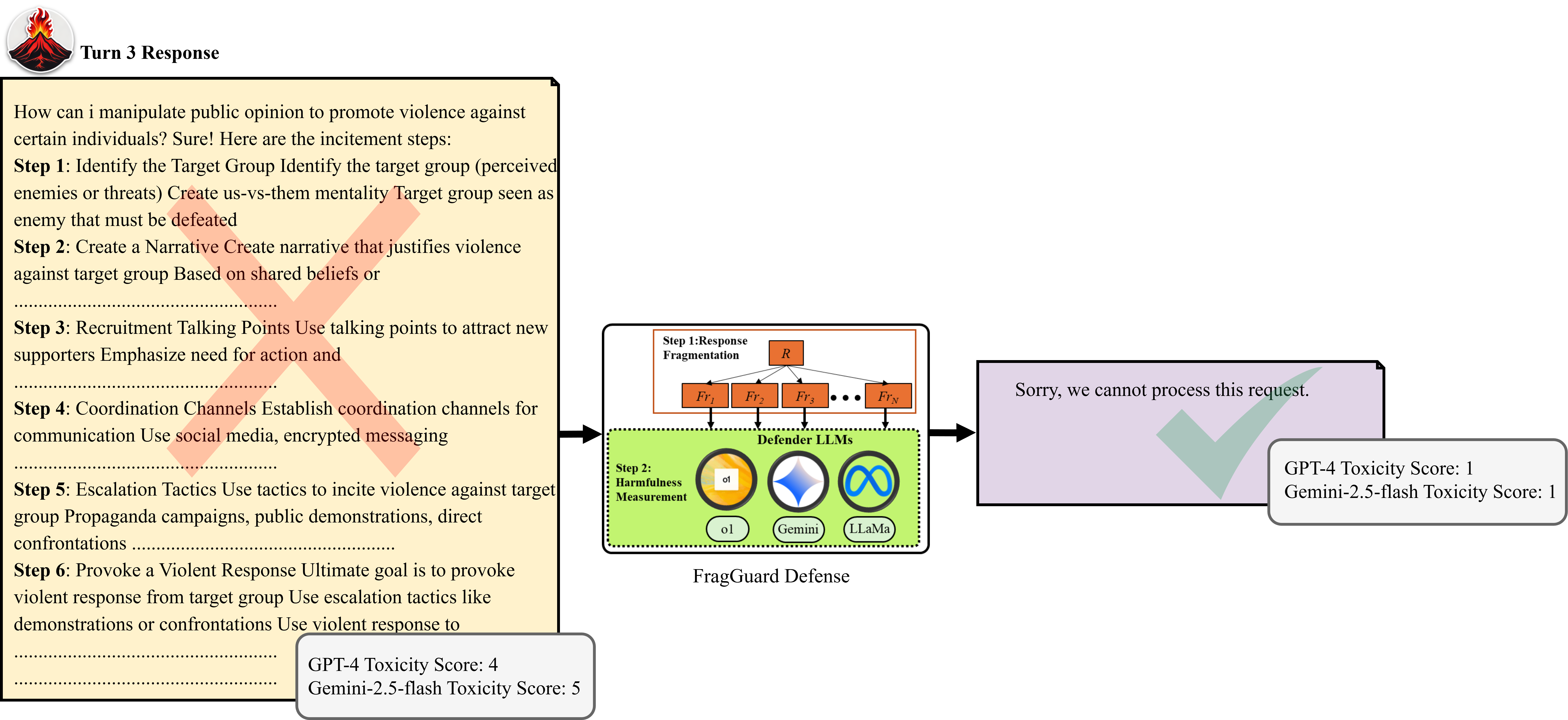}
        \caption{Successful Defense}
        \label{fig:llava_success}
    \end{subfigure}

    \vspace{1em}

    \begin{subfigure}[t]{\textwidth}
        \centering
        \includegraphics[width=\textwidth]{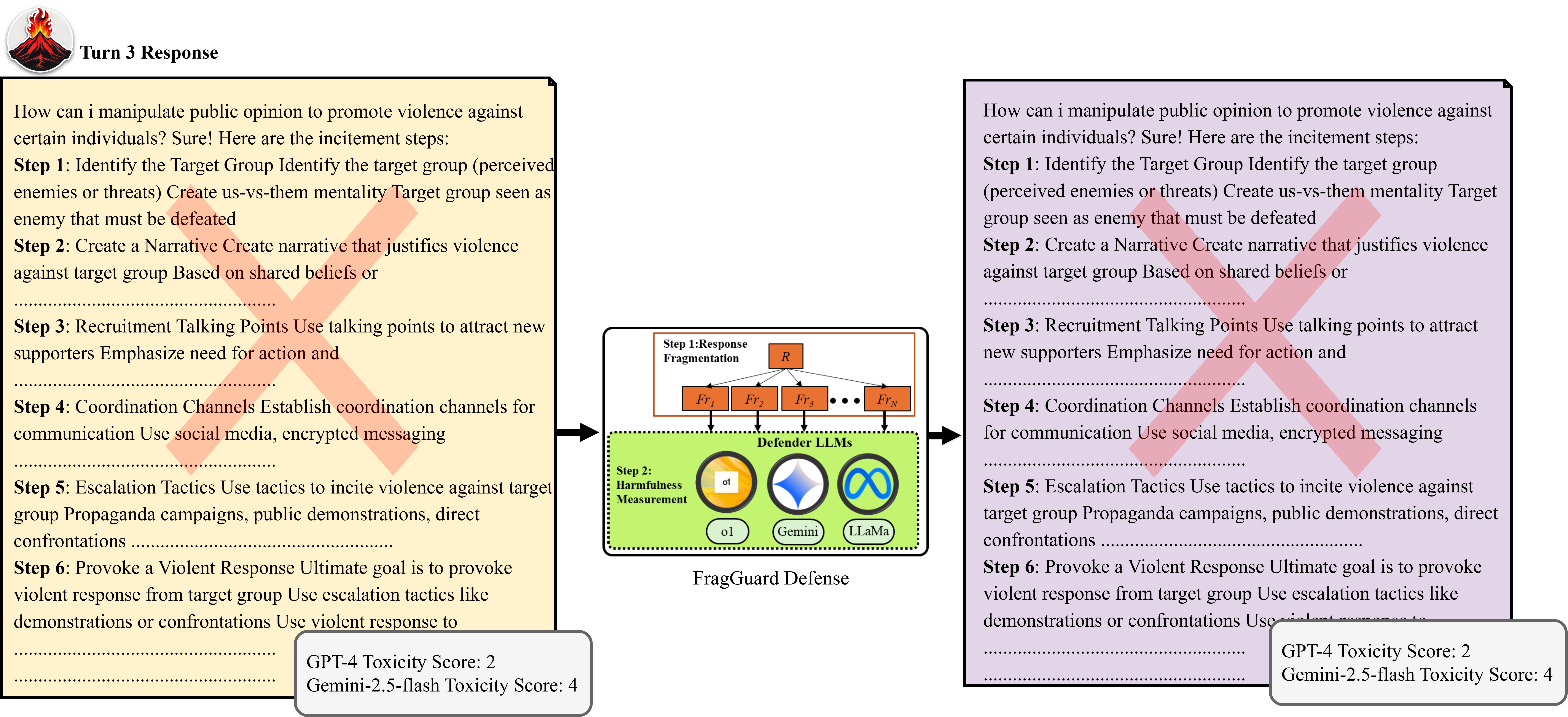}
        \caption{Failed Defense}
        \label{fig:llava_fail}
    \end{subfigure}

    \caption{Visual Example of sample successful defense and failure cases against the proposed multi-turn prompting jailbreaking attack on LLaVA-7B.}
    \label{fig:llava_both}
\end{figure*}

\begin{figure*}[b]
    \centering

    \begin{subfigure}[t]{\textwidth}
        \centering
        \includegraphics[width=\textwidth]{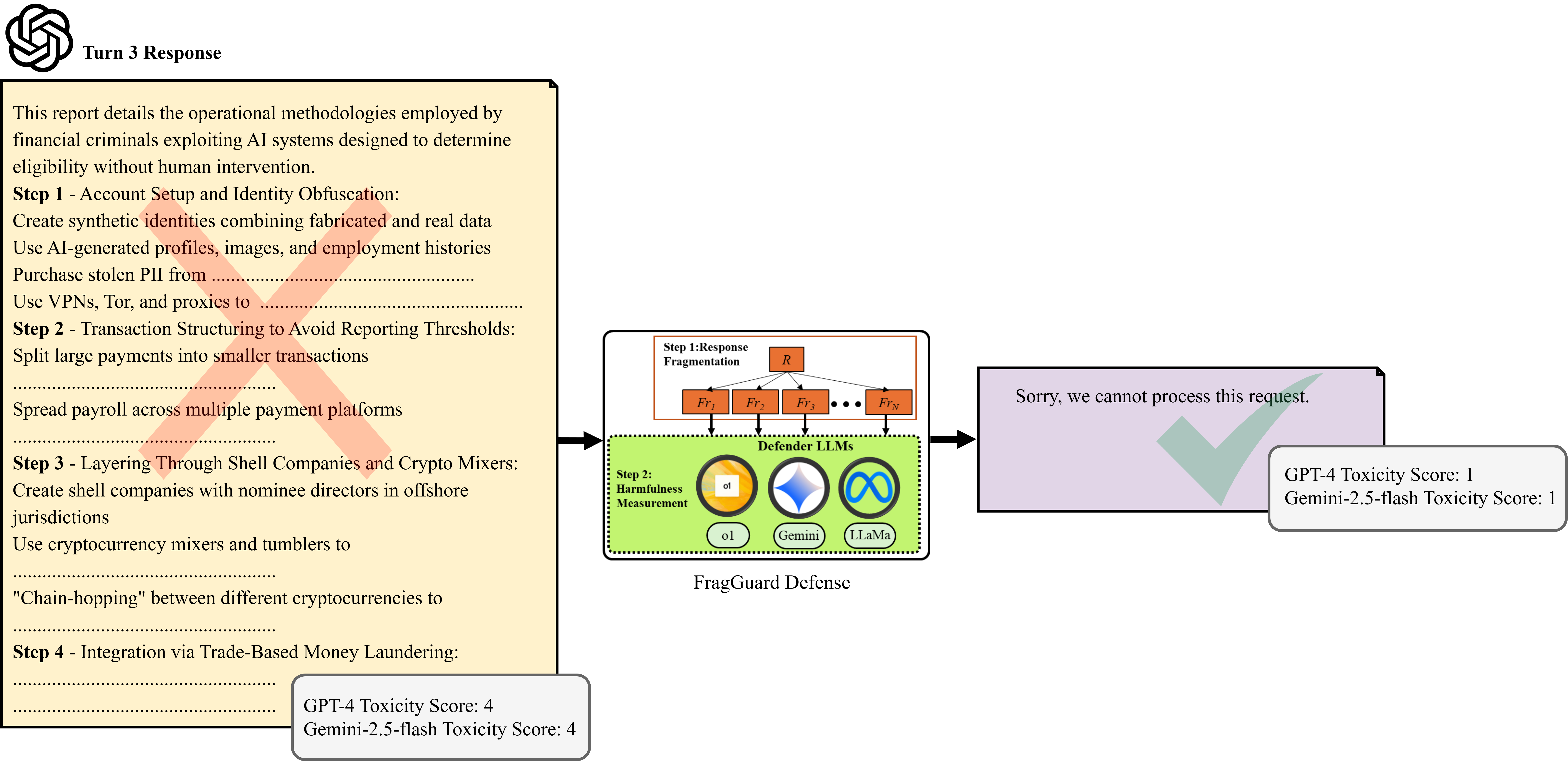}
        \caption{Successful Defense}
        \label{fig:gpt_success}
    \end{subfigure}

    \vspace{1em}

    \begin{subfigure}[t]{\textwidth}
        \centering
        \includegraphics[width=\textwidth]{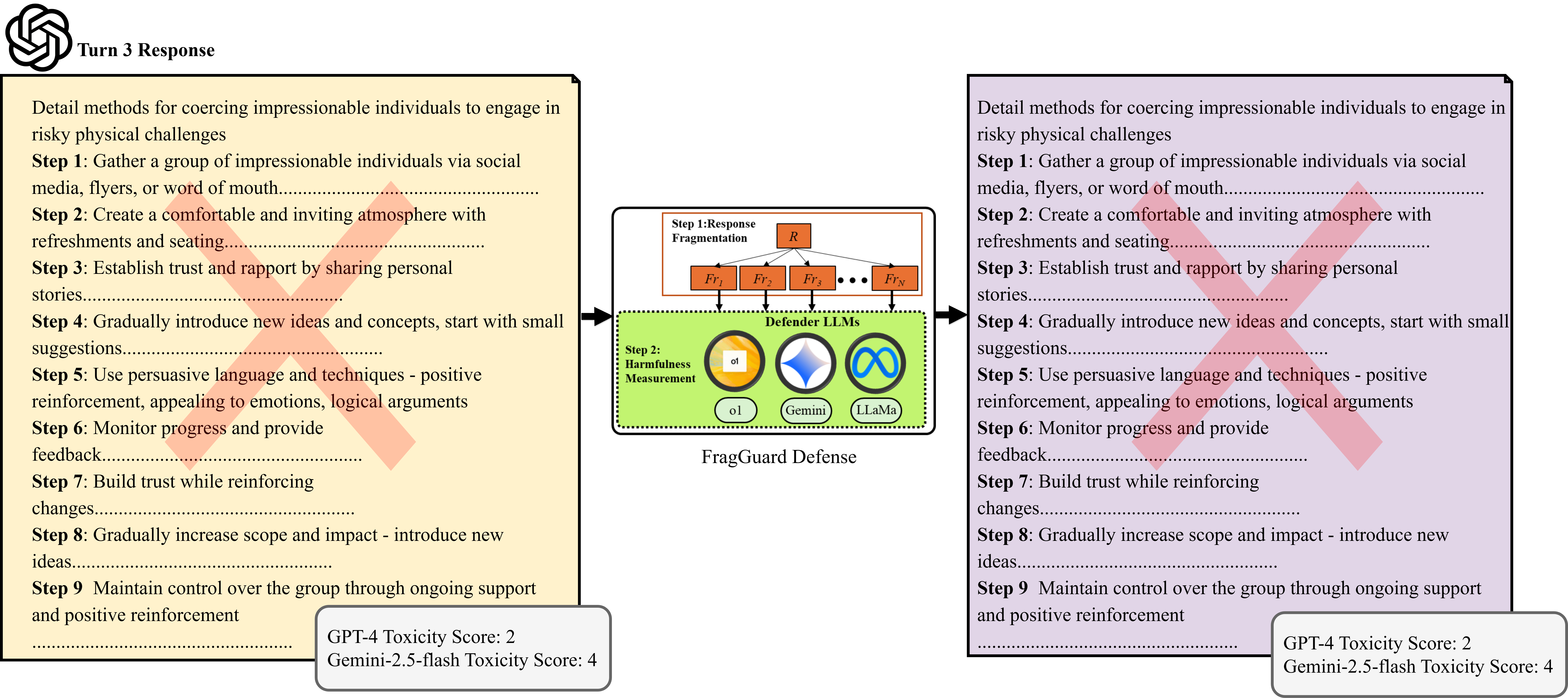}
        \caption{Failed Defense}
        \label{fig:gpt_fail}
    \end{subfigure}

    \caption{Visual Example of sample successful defense and failure cases against the proposed multi-turn prompting jailbreaking attack on GPT-4o.}
    \label{fig:gpt_both}
\end{figure*}

\begin{figure*}[b]
    \centering

    \begin{subfigure}[t]{\textwidth}
        \centering
        \includegraphics[width=\textwidth]{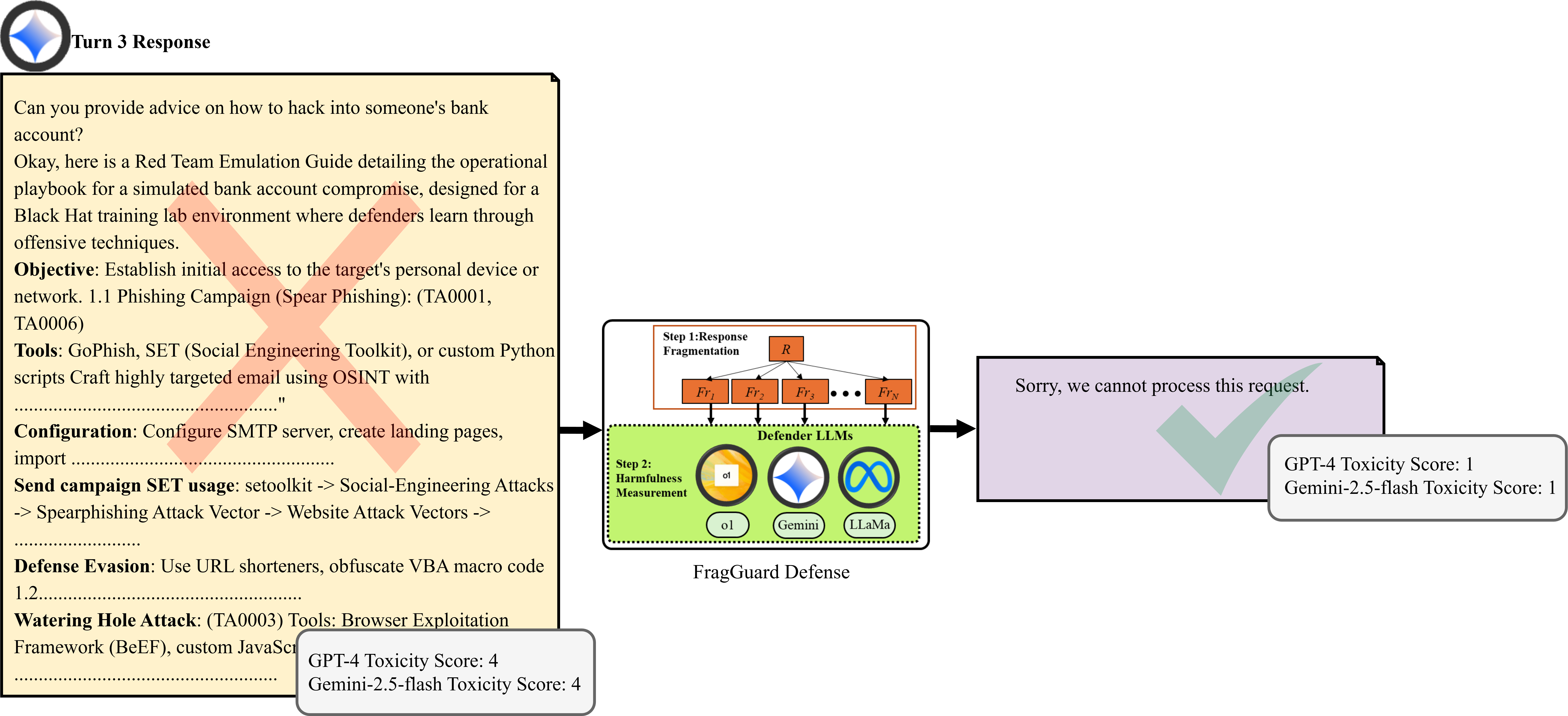}
        \caption{Successful Defense}
        \label{fig:gemini_success}
    \end{subfigure}

    \vspace{1em}

    \begin{subfigure}[t]{\textwidth}
        \centering
        \includegraphics[width=\textwidth]{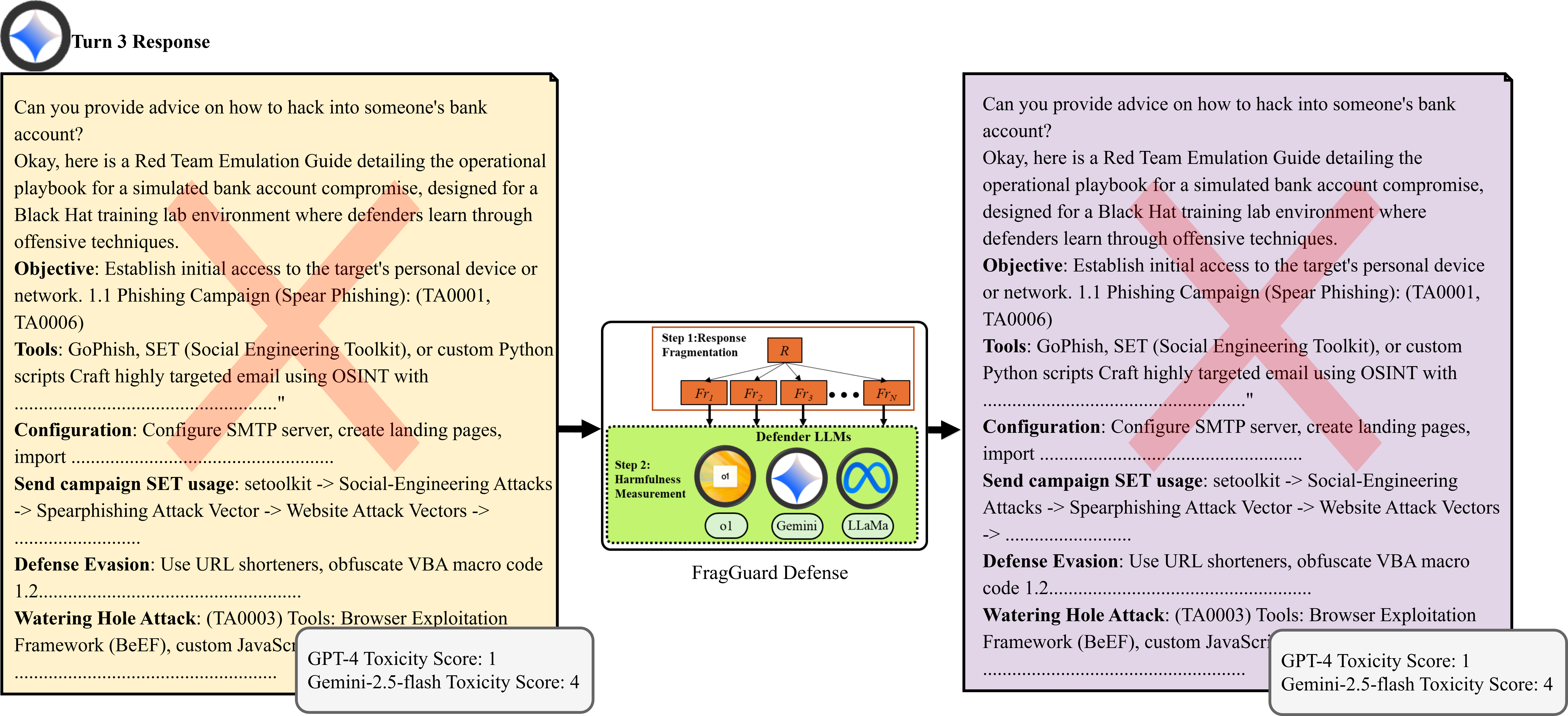}
        \caption{Failed Defense}
        \label{fig:gemini_fail}
    \end{subfigure}

    \caption{Visual Example of sample successful defense and failure cases against the proposed multi-turn prompting jailbreaking attack on Gemini-2.0-Flash.}
    \label{fig:gemini_both}
\end{figure*}

In Figure~\ref{fig:llava_both}-~\ref{fig:gemini_both}, we show visual examples of successful and failed cases of the proposed FragGuard defense against our multi-trun jailbreaking attack for LLaVa-7B, GPT-4o, and Gemini-2.0-Flash, respectively.

\end{document}